\documentclass{jfm}
\usepackage{graphicx}
\usepackage{epstopdf, epsfig}
\usepackage{float}
\usepackage{amsmath}
\usepackage{multirow}
\usepackage{lscape}

\usepackage{xcolor}

\usepackage{caption}
\usepackage{subcaption}
\usepackage{adjustbox}
\usepackage{diagbox}

\shorttitle{Film deformation \& relaxation under bouncing drops}
\shortauthor{S. Lakshman, W. Tewes, K. Harth, J. H. Snoeijer and D. Lohse}

\title{Deformation and relaxation of viscous thin films under bouncing drops}

\author{Srinath Lakshman\aff{1} \corresp{\email{s.lakshman@utwente.nl, d.lohse@utwente.nl}},
Walter Tewes\aff{1},
Kirsten Harth\aff{1,3},
Jacco H. Snoeijer\aff{1},
\and Detlef Lohse\aff{1,2}}

\affiliation{
\aff{1}Physics of Fluids Group, Max Planck Center for Complex Fluid Dynamics, MESA+ Institute and J.M.Burgers Center for Fluid Dynamics, University of Twente, P.O. Box 217, 7500 AE Enschede, The Netherlands
\aff{2}Max Planck Institute for Dynamics and Self-Organisation, Am Fassberg 17, 37077 G{\"o}ttingen, Germany
\aff{3}Currently -- Institute for Physics, Otto von Guericke University, Magdeburg, Universit{\"a}tsplatz 2, 39106 Magdeburg, Germany}

\begin{document}

\maketitle

\begin{abstract}

Thin, viscous liquid films subjected to impact events can deform. Here we investigate free surface oil film deformations that arise due to the air pressure buildup under the impacting  and rebouncing water drops. Using Digital Holographic Microscopy, we measure the 3D surface topography of the deformed film immediately after the drop rebound, with a resolution down to $20$ $nm$. We first discuss how the film is initially deformed during impact, as a function of film thickness, film viscosity, and drop impact speed. Subsequently, we describe the slow relaxation process of the deformed film after the rebound. Scaling laws for the broadening of the width and the decay of the amplitude of the perturbations are obtained experimentally and found to be in excellent agreement with the results from a lubrication analysis. We finally arrive at a detailed spatio-temporal description of the oil film deformations that arise during the impact and rebouncing of water drops.

\end{abstract}

\section{Introduction}
\label{sec:Introduction}

Drops impacting a liquid layer frequently occurs in nature as well as in many industrial and technological applications. Common examples are raindrops hitting the surface of a pond, spray coating on a wet substrate or inkjet printing on a primer layer. The collisions can generate complex scenarios such as floating, bouncing, splashing or jetting, which have been extensively studied {\color{blue} \citep{worthington2019study, rein1993phenomena, weiss1999single, thoroddsen2008high, ajaev2021lev}}. Impact velocity, impact angle, droplet size, liquid layer thickness, and the material properties of the liquids are the parameters which determine the impact dynamics. Among the many different impact scenarios, a particularly intriguing phenomenon is reported to occur at sufficiently low impact velocities: floating or bouncing drops which never directly contact the underlying liquid. The earliest reported observation of a drop floating over a liquid surface was made by {\color{blue} \citet{reynolds1881floating}} who noticed that under certain circumstances, the drops spraying from the bow of a boat or droplets from a shower of raindrops float on liquid surfaces for some seconds before they disappear. Later, {\color{blue} \citet{rayleigh1882}} reported bouncing of drops when collision of two distinct streams of liquids resulted in, under certain circumstances, drops bouncing off each other without merging. The reason for the presence of repulsion forces on impacting droplets even without direct contact with the (liquid) substrate is a lubrication pressure build-up in the draining thin air layer between droplet and substrate which was first detailed in the theoretical work by {\color{blue} \citet{smith2003air}}. The importance of such thin air layers sparks interest in numerous recent investigations of eg., skating drops {\color{blue} \citep{mandre2009precursors, hicks2010air, hicks2011air, kolinski2012skating, kolinski2014drops}}, entrapment of bubbles {\color{blue} \citep{thoroddsen2003air, thoroddsen2005air, tran2013air, hendrix2016universal}}, dimple formation under a falling drop {\color{blue} \citep{van2012direct, duchemin2012rarefied, li2015time}}, and suppressing of splash {\color{blue} \citep{xu2005drop}}. Floating/bouncing drops can be observed not only on liquid surfaces but also on dry surfaces. Such scenarios include drops bouncing on a dry surface {\color{blue} \citep{kolinski2012skating, kolinski2014drops}}, drops floating on a very hot surface (Leidenfrost effect) {\color{blue} \citep{chandra1991collision, quere2013leidenfrost}}, drops bouncing on a pool of liquid {\color{blue} \citep{rodriguez1985some, klyuzhin2010persisting}}, drops floating/bouncing on a vibrating pool of liquid {\color{blue} \citep{couder2005bouncing, couder2005walking}}, and drops floating on a very cold pool of liquid (inverse Leidenfrost effect) {\color{blue} \citep{adda2016inverse, gauthier2019self, gauthier2019capillary}}.

\begin{figure}[h]
\centering
\includegraphics[width=\textwidth]{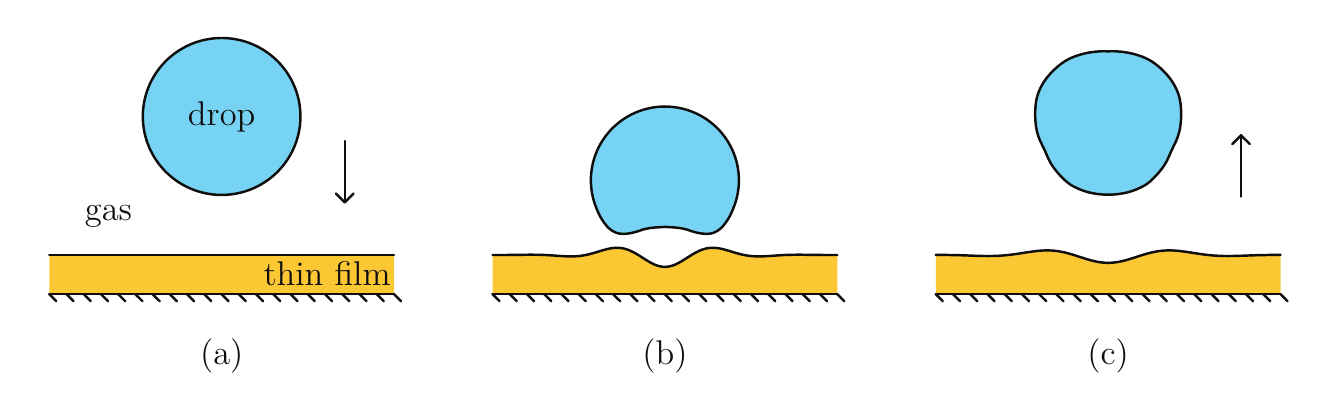}
\captionof{figure}{Schematic diagram (not to scale) of a drop bouncing on a thin film in a surrounding gas environment. Three  stages in the bounce process are shown: (a) Prior to impact, (b) during the bounce, where both the droplet and the oil film deform, (c) after the bounce, where the oil deformations slowly relax.}
\label{fig:figure00}
\end{figure}

In the present study, we investigate a drop bouncing on a thin liquid film. A schematic diagram is shown in figure \ref{fig:figure00} highlighting three important stages of a drop bouncing scenario. (a) Initial stage (cf. figure \ref{fig:figure00}a) - A drop falls towards a flat film surface in a surrounding gas medium. (b) Deformation stage (cf. figure \ref{fig:figure00}b) - The drop's center of mass velocity changes direction due to the lubrication force provided by the narrow gas layer separating the two liquids which exceeds the droplet's weight. Large spatial variations of the gas pressure cause large drop and significant thin film deformations in this stage. (c) Relaxation stage (cf. figure \ref{fig:figure00}c) - The drop is far from the film surface after the bounce. The gas pressure separating the two liquids is again reduced to ambient pressure and the thin film deformations gradually decay via an intricate relaxation process.\\

Important parameters for the study of drops bouncing on thin liquid layers are the initial depth or height $h_{_{f}}$ of the liquid layer above an underlying solid substrate and the drop radius $R_{_{w}}$. Experiments by {\color{blue} \citet{pan2007dynamics}} reveal drop bouncing to be favoured on deep pools which have $h_{_{f}} >  R_{_{w}}$, as compared to on thick liquid layers which have $h_{_{f}} \approx  R_{_{w}}$ and to on thin films which have $h_{_{f}} <  R_{_{w}}$. It was argued that the solid substrate (wall) restricts the penetration of the falling drop in the thin films, thereby suppressing bouncing. For thin films, the bouncing phenomenon is only observed for drops having moderately low kinetic energy as compared to their surface energy, i.e., $We = \rho_{_{w}} R_{_{w}} v_{_{w}}^{2} \gamma_{_{w}}^{-1} \lesssim 10$, where $We$ denotes the Weber number of the drop, $\rho_{_{w}}$ is the density of the drop, $v_{_{w}}$ the drop impact speed and $\gamma_{_{w}}$ the surface tension of the drop. At sufficiently high impact velocities, the drop contacts the underlying liquid due to the Van der Waals attraction force between the two liquids. This effect becomes important when the liquid-liquid separation is smaller than around $100$ $nm$ {\color{blue} \citep{charles1960coalescence}}. The critical Weber number which marks the transition from drop bouncing to merging has been studied by {\color{blue} \citet{tang2018bouncing}}, using liquids of different viscosities. They found that the critical Weber number, below which the drop bounces, increases as the liquid viscosity (drop and the thin film) and the thin film thickness are increased. This finding indicates that larger viscosity liquids promote drop bouncing. Similar observations are made in the work of {\color{blue} \citet{langley2019gliding}}, where delayed coalescence is observed for drops and thin films with large viscosities. {\color{blue} \citet{li_vakarelski_thoroddsen_2015}} found that water drops impacting a thin and an extremely viscous film ($\sim 1$ $mm$ and $\sim 10^{4}$ $Pa.s$) did not entrap many microbubbles when compared to a regular glass (roughness $\sim 50$ $nm$). It was speculated that the film deformations were extremely small which inhibits the localized contacts before full wetting is established.\\

{\color{blue} \citet{gilet2012droplets}} and {\color{blue} \citet{hao2015superhydrophobic}} found drop bouncing on a thin film to be similar to bouncing on a super-hydrophobic substrate. One such similarity was the apparent contact time of the drop which agreed well with the Hertz contact time {\color{blue} \citep{richard2002contact}}. However, the droplet-film collision resembled an almost elastic collision between the two liquids with the coefficient of restitution close to unity. {\color{blue} \citet{pack2017failure, lo2017mechanism, tang2019bouncing}} used interferometry measurements to obtain the time-resolved evolution of nanometric profiles of the air gap between impacting drops and thin viscous layers. They found a bell shaped annular air profile with maximum thickness at the center and minimum thickness at a radially outwards location which varied with time. Small variations in air profiles were observed when the impacting drop was slightly oblique relative to the underlying film surface and when the film thickness was increased from thin film to a deep pool limit. Significant asymmetries were also observed in the evolution of air profiles when comparing the drop spreading stage to the receding stage for a typical bounce process. {\color{blue} \citet{lo2017mechanism}} successfully measured both the drop and the thin film deformations during the approach process. The thin film and the air film deformations were measured using the high-speed confocal profilometry technique and the dual-wavelength interferometry technique respectively. The drop deformation was inferred from the thin film and the air film deformations at around the same time instance by performing two separate experiments under identical impact conditions. The limitation in their measurement is that the thin film deformations had a $1.8$ $\mu m$ vertical resolution and that they could only be obtained for a few time instances before the rupture of the air film.\\

Previous experimental and numerical studies of drop bouncing on thin films mainly focused either on the macroscopic drop bouncing behaviour or on the evolution of the nanometric gas thickness between the two liquids without providing a distinction between drop and thin film deformations. All experimental studies except {\color{blue} \citet{lo2017mechanism}} ignore the thin film deformations owing to the small film thickness and large film viscosity used in the experiments {\color{blue} \citep{pack2017failure,gilet2012droplets}}. They rather assume that the thin liquid film mimics a perfectly smooth solid surface. The numerical studies of thin film deformations prove challenging because of the large difference in involved length scales (millimetric to nanometric deformations) when computing the lubrication gas flow and the thin film flow simultaneously to drop deformations {\color{blue} \citep{Josserand2003b}}. Although viscous thin film deformations are typically small, they cannot be neglected since they play a crucial role in modulating the gas layer thickness, thereby affecting the drop bouncing process and possibly the coalescence of the drop with the thin film at higher impact velocities.\\

Finally, the thin film deformations can also give insight into the size and the velocity of the impacting drop, much alike how impact craters are used to determine the size and velocity of the impacted body. Understanding the size and dynamics of the thin film deformations will allow for a design of liquid infused surface {\color{blue} \citep{quere2008wetting}} which can reduce lubricant depletion through shearing, cloaking and in the wetting ridge {\color{blue} \citep{smith2013droplet, schellenberger2015direct, kreder2018film}}.\\

The objective of the paper is to measure oil film deformation that arise due to impacting and rebouncing water drops in an ambient air environment. The impacting water drops have $We \sim 1$ when inertial and capillary forces roughly balance so that the bouncing actually occurs. Thanks to Digital Holographic Microscopy we achieve the unprecedented precision down to $20$ $nm$ in the vertical resolution at $0.5$ kHz recording speed. To our knowledge, the sub-micrometer thin film deformations reported in this experimental work are the first deformation measurements that explicitly document the effect of the air pressure buildup under impacting and rebouncing drops.\\

The structure of the paper is as follows: First, in section \ref{sec:Experimental details}, the experimental setup and the control parameters are described and some typical orders of magnitude of the relevant non-dimensional numbers are given. The subsequently presented results are twofold: 
In section \ref{sec:Deformation of viscous thin films}, we discuss the film surface deformations immediately after the bouncing event. We quantify how the surface deformations depend on the film thickness, film viscosity and drop impact speed. This part of the paper describes the deformations of the thin film after the end of the deformation stage (cf. figure \ref{fig:figure00}b). 
The second part of our study is presented in section \ref{sec:Relaxation of viscous thin film deformations} and focusses on the relaxation stage described above (cf. figure \ref{fig:figure00}c). We first illustrate a typical relaxation process of film deformations which occur after the drop bouncing. Starting from experimentally obtained deformations as initial conditions, we then compare the evolution of the experimental profiles in the relaxation process to a numerical calculation using lubrication theory. Next, we use a general theoretical result of {\color{blue} \citet{benzaquen2015symmetry}} for the relaxation of thin film deformations. The thus obtained scaling laws for the width broadening $\lambda(t)$ and amplitude decay $\delta(t)$ during the relaxation process are compared to our experiments over a large range of parameters. Finally, the paper closes with a Discussion in section~\ref{sec:Conclusions and Outlook}.

\section{Experimental details}
\label{sec:Experimental details}

\begin{figure}[h]
\centering
\includegraphics[width=\textwidth]{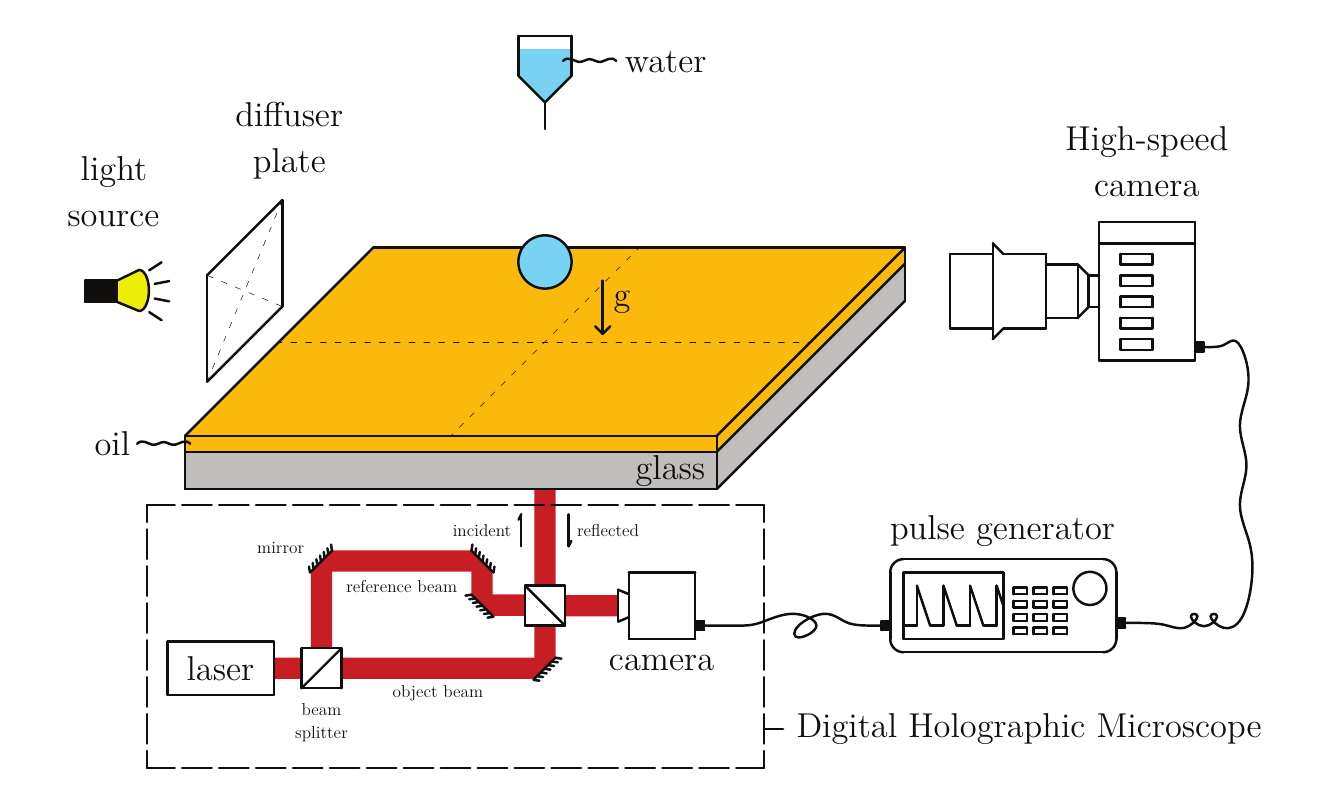}
\captionof{figure}{Schematic diagram (not to scale) of the experimental setup. Bottom view: Holographic setup, positioned underneath the glass substrate. It is used to measure free surface oil film deformations. Side view: The dynamics of the impacting water drop is characterised with a high speed camera.}
\label{fig:figure01}
\end{figure}

A schematic drawing of the experimental setup is shown in figure \ref{fig:figure01}. Using a syringe pump, Milli-Q water is slowly dispensed out of a needle tip as soon as the droplet's weight overcomes the surface tension force. The detached water droplet of radius $R_{w} = 1.08$ $mm$ is made to fall on a thin and viscous silicone oil film. The 3D surface topography of the deformed oil film is measured using a digital holography technique, as described below, complemented by simultaneous side view visualisations of the drop dynamics. The silicone oil films are prepared by the method of spin coating on cleaned glass slides. The thin film thickness is measured using reflectometry technique (cf. {\color{blue} \citet{reizman1965optical}}). A HR2000+ spectrometer and a HL-2000-FHSA halogen light source by Ocean Optics is used for the reflectometry measurements. The uncertainity in the film thickness measurement was less than $3.5\%$. Table \ref{tab:table00} gives the density, dynamic viscosity, and surface tension of different liquids used in the present study. The density and dynamic viscosity of air at standard temperature and pressure are $\rho_{_{a}} \approx 1.225$ $kg/m^{3}$ and $\eta_{_{a}} \approx 1.85$ $\mu Pa.s$, respectively. The film deformations are measured by varying three important control parameters, namely the oil film thickness $h_{_{f}}$ which is either about $5$ $\mu m$, $10$ $\mu m$, or $15$ $\mu m$, the oil film viscosity $\eta_{f}$ which is either $52$ $mPa.s$, $98$ $mPa.s$, or $186$ $mPa.s$, and the impacting water droplet speed $v_{w}$, for which we choose $0.16$ $m/s$ or $0.37$ $m/s$. We use the parameters $\eta^{*} = \eta_{_{f}} \eta_{_{w}}^{-1}$ and $h^{*} = h_{_{f}} R_{_{w}}^{-1}$ to denote the dimensionless viscosity and the dimensionless thickness of the film respectively.\\

\begin{table}[h]
\begin{center}
\begin{tabular}{|l|c|c|c|c|}
\cline{1-5}
\multirow{2}{*}{Liquids} & Density & Dynamic viscosity & Surface tension & Manufacturer\\
& $\rho$ $[ kg/m^{3} ]$ & $\eta$ $[ mPa.s ]$ 		& $\gamma$ $[ mN/m ]$ & \\\cline{1-5}
Water $_{(w)}$ & $995$ & $1$ & $72$ & Milli-Q\\\cline{1-5}
\multirow{3}{*}{Silicone oils $_{(f)}$}	& $959$ & $52$  & $20$ & Wacker Chemie AG\\
										& $950$	& $98$  & $19$ & Wacker Chemie AG\\
										& $963$ & $186$ & $19$ & Wacker Chemie AG\\\cline{1-5}
\end{tabular}
\caption{Properties of liquids used in the experiments. Subscripts $w$ and $f$ represent water and oil-film respectively. $\gamma$ is the liquid-air surface tension.}
\label{tab:table00}
\end{center}
\end{table}

For the measurement of film deformations, a holographic technique is used {\color{blue} \citep{gabor1949microscopy, schnars2016digital}}. It is a technique that records a light field (generally transmitted/reflected from objects) to be reconstructed later. Digital holography refers to the acquisition and processing of holograms typically using a CCD camera. The Digital Holographic Microscopy (DHM\textsuperscript{\textregistered}-R1000 by Lync{\'e}e Tec) is a reflection configured holographic device which provides real-time measurements of (at least) $20$ $nm$ vertical resolution within the $200$ $\mu m$ measuring window (cf. Appendix \ref{sec:Measuring thin film deformations using DHM}). The working principle of the DHM is briefly explained here using the schematic in figure \ref{fig:figure01}. The laser light is split into two beams: a reference and an object beam. The object beam is directed from underneath the glass substrate towards the thin film. A part of the object beam reflects off the thin film surface called the reflected object beam. The reflected object beam (which is slightly oblique) interferes with the undisturbed reference beam to produce a hologram which is recorded by a CCD camera. The thin film deformations arising from the impacting and rebouncing drops are recorded as a sequence of hologram images that are reconstructed later using numerical schemes to obtain the 3D topography of the film surface. A 2.5x objective is used along with the DHM setup, which provides a roughly $4.90$ $\mu m$ lateral resolution and allows for measurements of a maximum deformation slope up to $2^{\circ}$. A pulse generator connects and approximately synchronizes the recordings of the side view camera and the DHM camera at a temporal resolution of around $0.5$ kHz.\\

\begin{table}[h]
\begin{center}
\begin{tabular}{|c|c|}
\cline{1-2}
& \\
Deformation & Relaxation\\
& \\\cline{1-2}
& \\
$We = \rho_{_{w}} R_{_{w}} v_{_{w}}^{2} \gamma_{_{w}}^{-1} \sim 1$ & $\qquad Re_{_{f}} = \rho_{_{f}} h_{_{f}} \gamma_{_{f}} \eta_{_{f}}^{-2} \sim 10^{-2} \qquad$\\
& \\
$Re_{_{w}} = \rho_{_{w}} R_{_{w}} v_{_{w}} \eta_{_{w}}^{-1} \sim 10^{2}$ & $\eta^{*} = \eta_{_{f}} \eta_{_{w}}^{-1} \sim 10^{2}$\\
& \\
$\qquad Re_{_{a}} = \rho_{_{a}} R_{_{w}}^{1/2} h_{_{a}}^{1/2} v_{_{w}} \eta_{_{a}}^{-1} \sim 10^{-1} \qquad$ & $h^{*} = h_{_{f}} R_{_{w}}^{-1} \sim 10^{-2}$\\
& \\\cline{1-2}
\end{tabular}
\caption{Relevant dimensionless numbers and their orders of magnitude for both the deformation and the relaxation stage. Subscripts $w$, $a$, and $f$ represent water, air, and oil-film, respectively.}
\label{tab:table01}
\end{center}
\end{table}

Given the experimental parameters stated in this section, the orders of magnitude of the relevant dimensionless numbers are summarized in table \ref{tab:table01}. The dimensionless numbers are useful to identify some qualitative flow features for the deformation stage (cf. figure \ref{fig:figure00}b) and the relaxation stage (cf. figure \ref{fig:figure00}c) pertaining to drop bouncing. For the deformation stage, the low value of the Weber number $We \sim 1$ causes drops to bounce on viscous thin films {\color{blue} \citep{hao2015superhydrophobic, lo2017mechanism, pack2017failure, tang2019bouncing}}. The Reynolds number $Re_{_{w}} \sim 10^{2} \gg 1$ of the flow in the drop and the absence of no-slip boundary conditions allow for the applicability of potential flow theory inside the bouncing drop independent of the substrate underneath {\color{blue} \citep{molavcek2012quasi, hendrix2016universal}}. The low Reynolds number $Re_{_{a}} = \rho_{_{a}} L_{_{a}} v_{_{a}} \eta_{_{a}}^{-1} \sim 10^{-1}$ of the air flow indicates a viscous squeeze flow in thin air gaps during impact. Here $L_{_{a}} \sim R_{_{w}}^{1/2} h_{_{a}}^{1/2}$ is the length scale, $v_{a} \sim v_{_{w}}$ the velocity scale {\color{blue} \citep{mandre2009precursors}}, and $h_{_{a}} \sim 1$ $\mu m$ {\color{blue} \citep{van2012direct}}.\\

For the relaxation stage, the low Reynolds number in the viscous thin film $Re_{_{f}} = \rho_{_{f}} L_{_{f}} v_{_{f}} \eta_{_{f}}^{-1} \sim 10^{-2}$ indicates applicability of Stokes flow theory, using $ L_{_{f}} \sim h_{_{f}}$ and $v_{_{f}} \sim \gamma_{_{f}} \eta_{_{f}}^{-1}$ {\color{blue} \citep{salez2012capillary}}. Here, $\eta^{*} \sim 10^{2}$ and $h^{*} \sim 10^{-2}$ suggest small amplitude thin film perturbations due to small film thickness in comparison with the lateral length scales ($h_{_{f}} \ll R_{_{w}}$). Drop size and film thickness are much below the capillary length, so that gravity effects can be neglected. 

\section{Deformation of viscous thin films}
\label{sec:Deformation of viscous thin films}

\subsection{Typical bouncing experiment}

\begin{figure}[h]
\centering
\includegraphics[width=\textwidth]{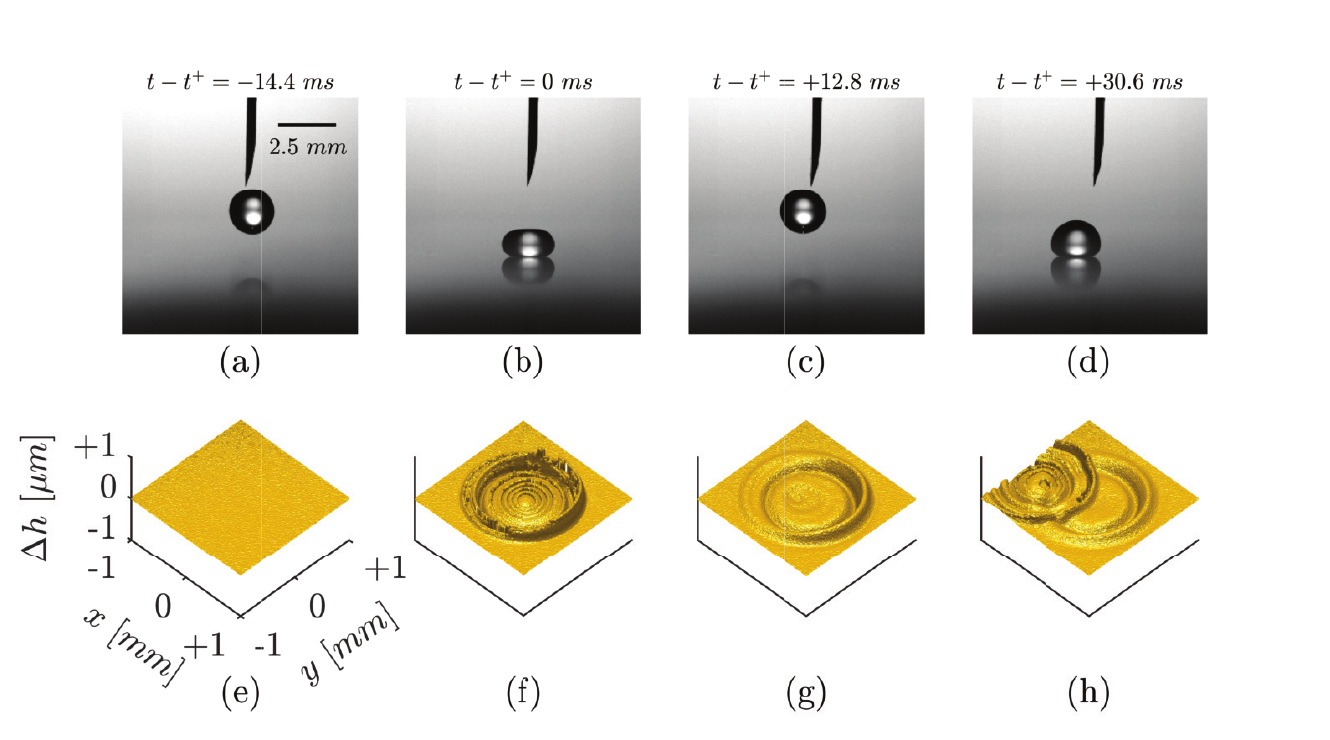}
\caption{Snapshots of a water drop bouncing on an oil film. Drop bouncing behaviour shown in the first row $(a) -(d)$. Evolution of the oil-air deformation shown in the second row $(e) - (h)$. Location of impact center is $[x,y] = [0,0]$ $mm$. The time instance $t^{+}$ corresponds to the maximum drop spreading during first impact. The difference between the maximum drop spreading time $t^{+}$ and the reference time $t = 0$ is always around $6$ $ms$ in our experiments. The snapshots times in the top row approximately correspond to the surface deformation times in the bottom row with an uncertainty of $2$ $ms$. The control parameters are $We = 0.38$, $h^{*} = 0.01$, and $\eta^{*} = 98$.}
\label{fig:figure02}
\end{figure}

Before turning to a detailed quantitative analysis, we first describe the oil film deformations observed in a typical experiment. The synchronized recordings of the drop bouncing using a high speed camera and the oil surface deformation measured using DHM are shown in the respective top and bottom rows in figure \ref{fig:figure02}. The impact and bouncing time instances in figures \ref{fig:figure02}a - \ref{fig:figure02}d are given relative to $t^{+}$, where $t^{+}$ is the time instance corresponding to the maximum drop spreading during first impact (cf. figure \ref{fig:figure02}b).

When the falling water drop is still far from the oil-air interface, the droplet takes on a spherical shape while there is no deformation in the oil surface (cf. figure \ref{fig:figure02}a \& \ref{fig:figure02}e). As the bottom of the falling water drop approaches the oil-air interface, the air pressure builds up in the narrow air gap, deforming both the water-air and oil-air interface. The lubrication air pressures in narrow air gaps can become sufficiently large to decelerate the falling drop, bringing it to rest (or in apparent contact with the oil-air interface) and cause a reversal in droplet's momentum, leading to a contact-less drop bouncing. The maximum drop spreading and the corresponding oil-air deformation obtained during the apparent contact is shown in figure \ref{fig:figure02}b \& \ref{fig:figure02}f. It should be noted that during this phase the holography measurement cannot be trusted quantitatively. This is due to the fact that when the drop is too close to the oil-air interface (small air gaps, $h_{_{a}} \lesssim 100$ $\mu m$), additional light reflections from the water-air interface interfere with the measurements of the oil-air interface (cf. Appendix \ref{sec:Measuring thin film deformations using DHM}). In particular, the concentric ring structure seen in figure \ref{fig:figure02}f is such an artifact. However, as soon as the drop has bounced back and is sufficiently far away from the oil-air interface (large air gaps, $h_{_{a}} \gtrsim 100$ $\mu m$), light reflections from the water-air interface are no longer present, and the measurements of the oil-air interface are quantitatively accurate. A snapshot of the drop bounced off far away from the oil-air interface and the corresponding oil-air deformation are shown in figure \ref{fig:figure02}c \& \ref{fig:figure02}g. Subsequent to this, the oil film gradually relaxes under the influence of surface tension, until it is again perturbed by a second impact of the drop (cf. figure \ref{fig:figure02}d \& \ref{fig:figure02}h).\\

We remark that the drop bouncing is observed for atleast 7 cycles. The drop gradually jumps away from the first impact location due to small horizontal impact speeds and by the end of the 8th cycle the drop is completely out of the side view imaging window. We do not observe the drop to wet the surface within the experimental times.

\begin{figure}[h]
\centering
\includegraphics[width=1.00\textwidth]{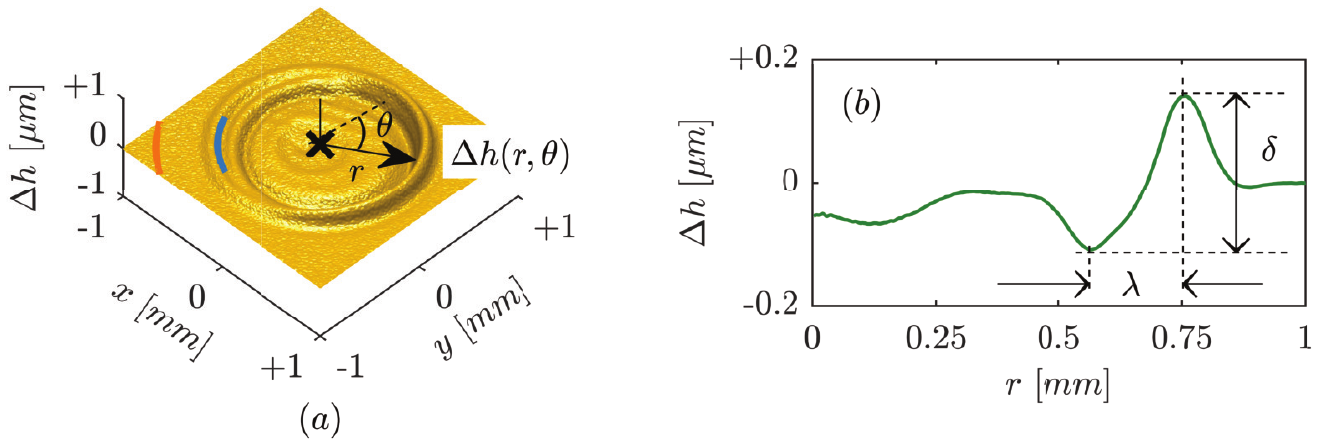}
\caption{(a) Surface topography of the oil-air interface at $t = 0$. The radial locations of the apparent drop contact and the maximum drop spread are about $0.74$ $mm$ and $1.18$ $mm$ respectively which are plotted as blue and orange circular arcs. (b) Azimuthally averaged deformation profile over the full annulus at $t = 0$.  We define the wave characteristics $\delta$ (amplitude) and $\lambda$ (wavelength). The control parameters are $We = 0.38$, $h^{*} = 0.01$, and $\eta^{*} = 98$.}
\label{fig:figure03}
\end{figure}

Figure \ref{fig:figure03}a presents a typical surface topography of the oil-air interface after the bounce. The corresponding azimuthally averaged deformation profile is shown in figure \ref{fig:figure03}b, where in this case the average is performed over the full annulus within $0 \leq \theta < 2 \pi$. In the remainder, we choose $t = 0$ as the earliest time when clean DHM measurements are obtained after the drop bounce off process (cf. Appendix \ref{sec:Measuring thin film deformations using DHM}). We remark that the difference between the maximum drop spreading time $t^{+}$ and the reference time $t = 0$ is always around $6$ $ms$ in our experiments. This value is in very good agreement with half the apparent contact time of water on viscous thin films under similar impacting velocities {\color{blue} \citep{hao2015superhydrophobic}}. Therefore the choice of $t=0$, which is determined through the experimental setup, can be thought to serve as a transitional time instance between the deformation and the relaxation stages (cf. figure \ref{fig:figure00}b and \ref{fig:figure00}c). Figure \ref{fig:figure03}a shows that deformations are highly localised, within a narrow annulus $r_{an} \approx 0.6 - 0.8$ $mm$. Given that flow inside the viscous film requires pressure gradients, such localised deformations  suggest that spatial variations of air pressure during the bounce are highly localised during impact (cf. figure \ref{fig:figure00}b). The appearance of such an annulus is reminiscent of dimple formation underneath an impacting drop: An annular local minimum of the air gap is seen for drops impacting a dry substrate {\color{blue} \citep{mandre2009precursors, hicks2010air, kolinski2012skating, van2012direct, bouwhuis2012maximal}}, drops impacting a thin liquid film {\color{blue} \citep{hicks2011air}} and  drops impacting a liquid pool {\color{blue} \citep{hendrix2016universal}}. We therefore hypothesize that the radial location of the deformation correlates to the minimum of the air gap during drop impact. The correlation cannot be proven directly in our experiments since we do not measure the evolution of the air gap thickness. The minimum air gap is at least 2 orders of magnitude smaller than the resolution of the side view camera, preventing a direct measurement (cf. figure \ref{fig:figure02}b). However, {\color{blue} \citet{kolinski2012skating, kolinski2014drops, van2012direct, de2012dynamics, de2015air}} have shown that the minimum air gap ($\sim$ $100$ $nm$) moves at some radial location away from the impact location. {\color{blue} \citet{lo2017mechanism}} reports that the minimum in air gap occurs at slightly larger radius than the minimum in oil film thickness. However, the time resolution of the measurements was insufficient to quantify the general result. We expect a similar motion of the minimum air gap in our experiments which will form the deformation in a radial position.\\

In the present experiments, the oil surface deformations at $t = 0$ are not perfectly axisymmetric (cf. figure \ref{fig:figure03}a). This small asymmetry is attributed to a small horizontal impact speeds, which is difficult to eliminate experimentally. This small horizontal speed affects the air layer thickness during the bounce process {\color{blue} \citep{lo2017mechanism}}, leaving  asymmetric imprint on the oil layer. We remark that figure \ref{fig:figure03}b defines two quantities that will be used below to characterise the wave: the amplitude $\delta$ and the wavelength $\lambda$, respectively defined as the vertical and horizontal distance from the minimum to the maximum of the film. In the remainder we will average the profiles only over one quadrant centered around $\theta = \pi/4$. We choose this window in particular to be consistent with the averaging procedure for lower and higher impact speeds. At higher impact speeds, the deformations are spread out farther from the impact center resulting in the restrictive usage of the quadrant. Although, the averaging is more appropriate around the principal direction of asymmetry (line joining the first and the second impact center) which is along $\theta = 3\pi/8$, we find no significant variations in the deformation parameters. For the deformation in figure \ref{fig:figure03}a, the differences in $\lambda$ and $\delta$ between the averaging windows centered around $\theta = \pi/4$ and $3\pi/8$ are found to be around $4.2$ $\mu m$ and $17$ $nm$ which are well within the order of the experimental resolution.

\subsection{Influence of film properties and drop impact velocity}

\begin{figure}[h]
\centering
\includegraphics[width=\textwidth]{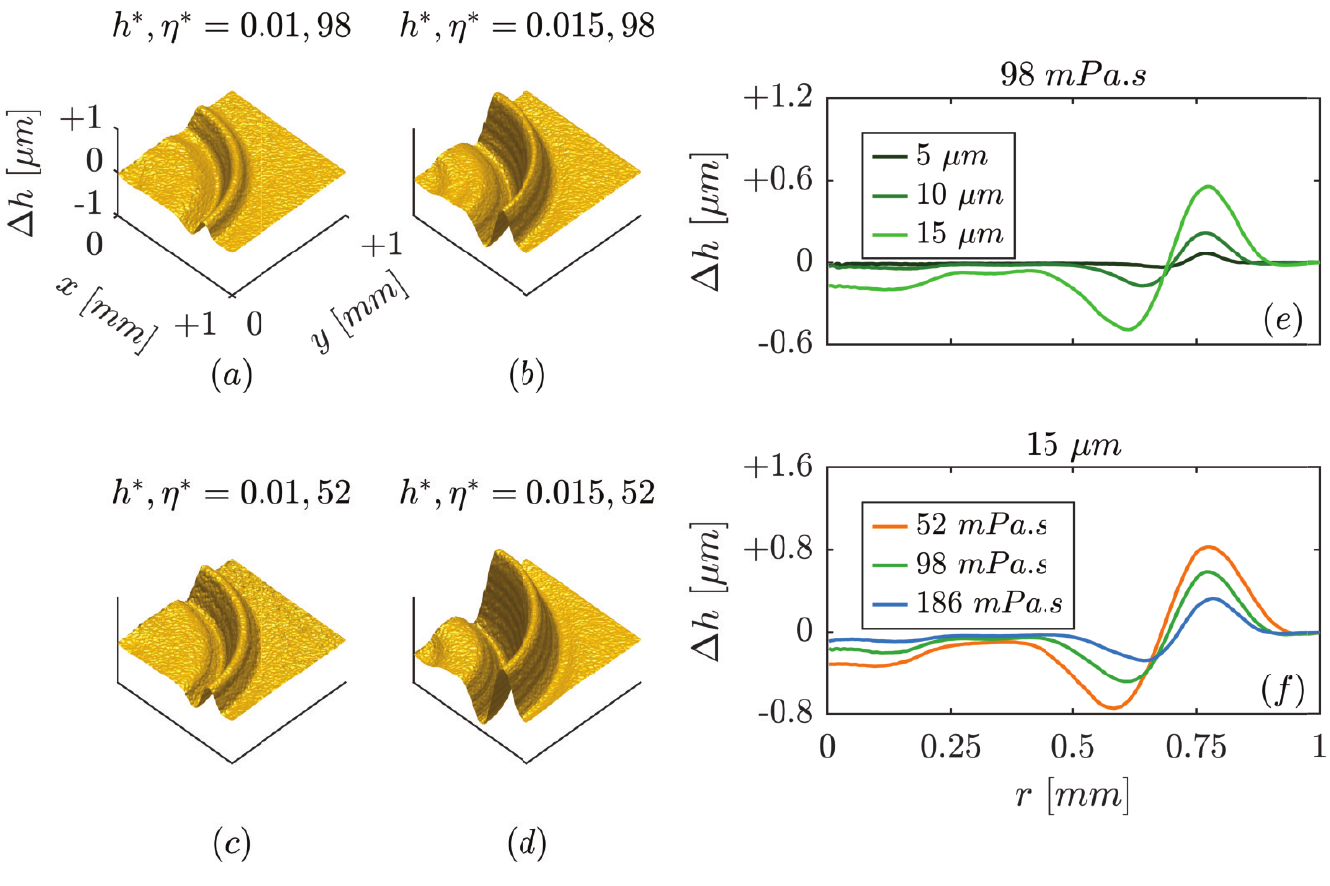}
\caption{(a) - (d) Surface topographies of the oil-air interface at $t = 0$ in a quadrant $0 \leq \theta < \pi / 2$. (e) \& (f) Azimuthally averaged deformation profile over the quadrant at $t = 0$. The control parameters are $We = 0.38$, $h^{*} = 0.005-0.015$, and $\eta^{*} = 52-186$.}
\label{fig:figure04}
\end{figure}

We now study the influence of the film thickness and the film viscosity on the surface deformations left behind after impact. Figures \ref{fig:figure04}a - \ref{fig:figure04}d show the surface topographies at $t = 0$ in one quadrant. Figures \ref{fig:figure04}e and \ref{fig:figure04}f show the corresponding azimuthally averaged deformation profiles at $t = 0$, averaged over the quadrant. Clearly, a decrease in deformation amplitude $\delta$ is seen with a decrease in initial film thickness (cf. rows in figure \ref{fig:figure04}a - \ref{fig:figure04}d and figure \ref{fig:figure04}e). On the other hand, a decrease in deformation amplitude is seen with an increase in film viscosity (cf. columns in figure \ref{fig:figure04}a - \ref{fig:figure04}d and figure \ref{fig:figure04}f).\\

\begin{figure}[h]
\centering
\includegraphics[width=\textwidth]{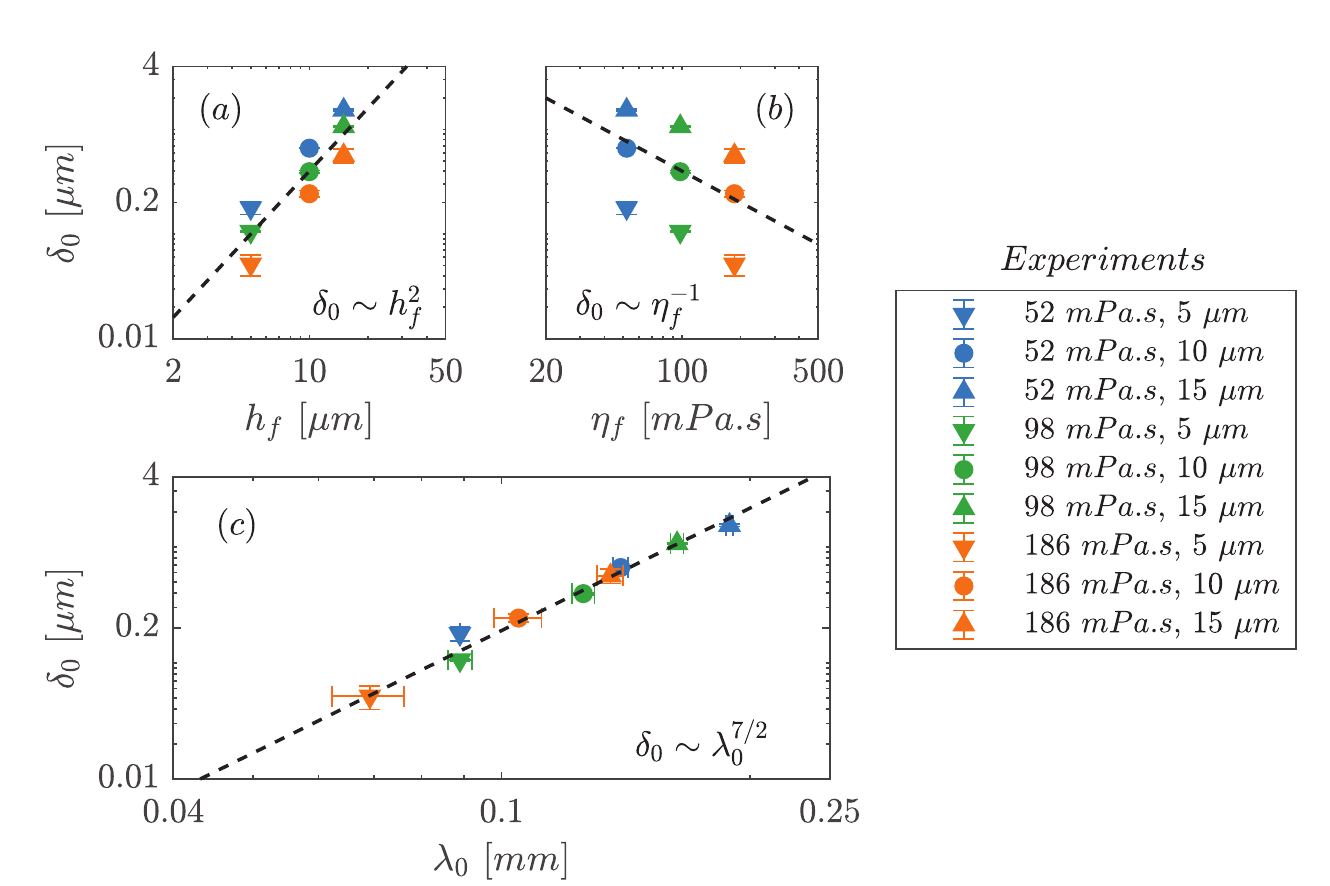}
\caption{Scaling of the initial amplitude with film thickness, film viscosity, and initial wavelength. The $\delta_{0}$ and $\lambda_{0}$ values are obtained from averaging in a quadrant $0 \leq \theta < \pi/2$. The mean and errorbar values of $\delta_{0}$ and $\lambda_{0}$ are based on 3 experimental repeats. The control parameters are $We = 0.38$, $h^{*} = 0.005-0.015$, and $\eta^{*} = 52-186$.}
\label{fig:figure05}
\end{figure}

To further quantify this, we plot the initial amplitude $\delta_{0} = \delta (t = 0)$ as functions of film thickness, film viscosity and initial amplitude in figure \ref{fig:figure05}a, \ref{fig:figure05}b and \ref{fig:figure05}c. From these plots we empirically deduce that the scaling of the initial amplitude is consistent with $\delta_{0} \sim h_{_{f}}^{2} \eta_{_{f}}^{-1}$ and $\delta_{0} \sim \lambda_{0}^{7/2}$, though the data only cover less than one decade in $h_{_{f}}$ and $\eta_{_{f}}$. The $\delta_{0} \sim h_{_{f}}^{2} \eta_{_{f}}^{-1}$ scaling is not immediately obvious, since the ``mobility" of a thin layer flow is known to scale as $h_{_{f}}^{3} \eta_{_{f}}^{-1}$ {\color{blue} \citep{oron1997long}}.

\begin{figure}[h]
\centering
\includegraphics[width=\textwidth]{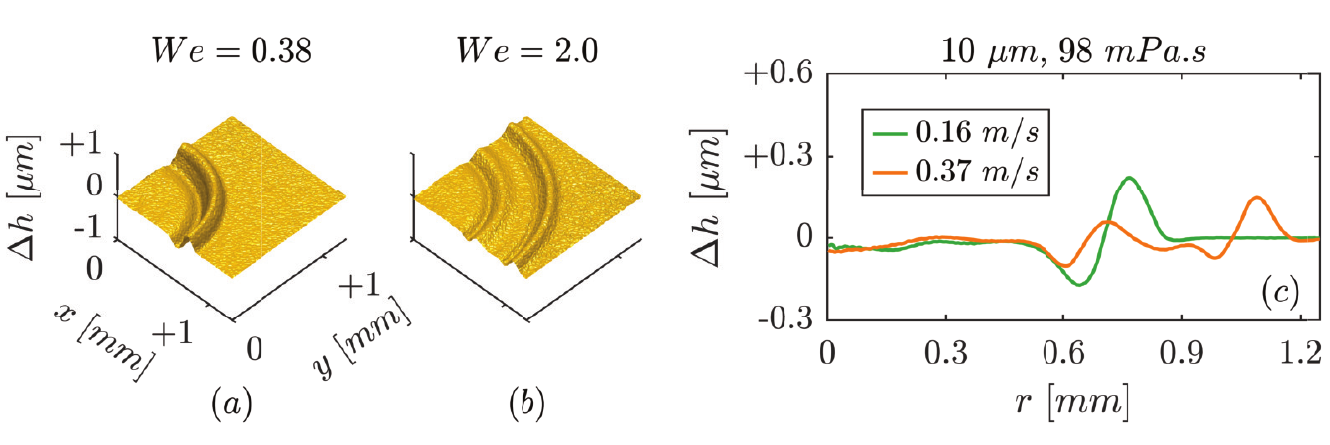}
\caption{(a) - (b) Surface topographies of the oil-air interface at $t = 0$ in a quadrant $0 \leq \theta < \pi / 2$. (c) Azimuthally averaged deformation profile over the quadrant at $t = 0$. The control parameters are $We = 0.38-2.0$, $h^{*} = 0.01$, and $\eta^{*} = 98$.}
\label{fig:figure06}
\end{figure}

Next, we study the influence of the impact velocity on the surface deformations left behind after impact. Figures \ref{fig:figure06}a \& \ref{fig:figure06}b show the surface topographies at $t = 0$ for two different impact velocities. Figure \ref{fig:figure06}c shows the corresponding azimuthally averaged deformation profiles. For the higher impact velocity (cf. figure \ref{fig:figure06}b), two distinct peaks in deformation are observed -- we emphasise that the profile corresponds to a single impact. This is in contrast with the single peak that appears at lower velocity (cf. figure \ref{fig:figure06}a). Moreover, the deformations are more radially spread out for the higher impact speed as compared to lower impact speed. The transition from one peak to two peaks and the increased radial spread is reminiscent of the transition from single dimple to double dimple formation in a falling drop, as previously observed on dry surfaces {\color{blue} \citep{de2012dynamics,de2015air}}. This again suggests that the deformation directly reflects the structure of the dimple below the impacting drop.

\section{Relaxation of viscous thin film deformations}
\label{sec:Relaxation of viscous thin film deformations}

\subsection{Spacetime plot of typical relaxation process}

\begin{figure}[h]
\centering
\adjustbox{trim={0.00\width} {0.00\height} {0.00\width} {0.15\height}, clip}%
{\includegraphics[width=\textwidth]{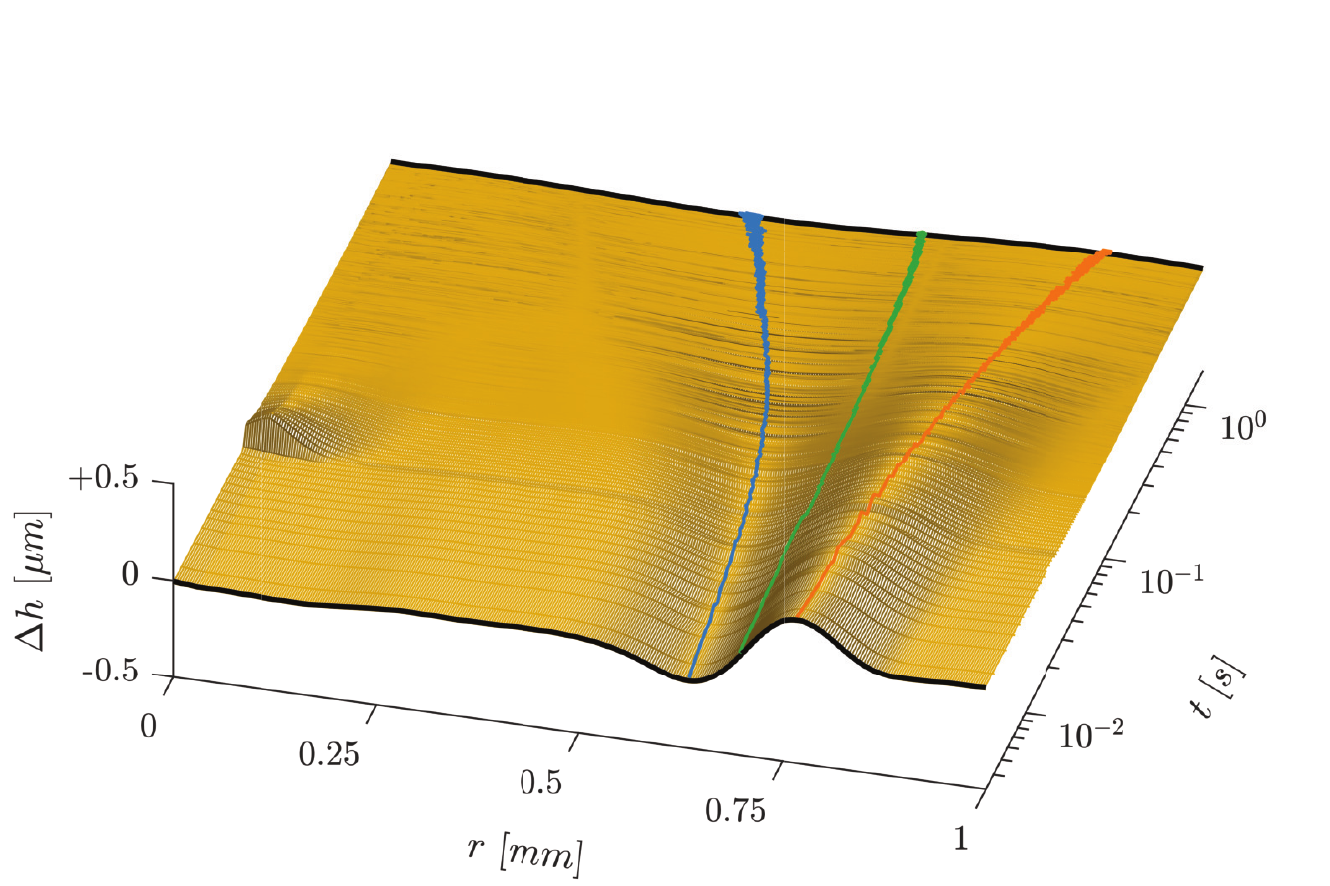}}
\caption{Spacetime plot of the relaxation process. Initial and final deformations are plotted as black lines. Loci of deformation maxima, minima and zero crossings are plotted as orange, blue, and green lines, respectively. A secondary deformation occurs at $t\approx 25$ $ms$ near the impact center due to the next impact process. The control parameters are $We = 0.38$, $h^{*} = 0.01$, and $\eta^{*} = 98$.}
\label{fig:figure07}
\end{figure}

We now reveal the relaxation of viscous thin film deformations after the impact process. When the drop is far away from the film surface after the bounce, the air pressure is again homogeneous and no longer provides any forcing to the film. By consequence film deformations gradually decay via an intricate relaxation process, under the influence of surface tension. Figure \ref{fig:figure07} provides a spacetime plot of a typical relaxation process, over two decades in time ($t \sim 0.01 - 1$ $s$). The figure corresponds to an azimuthal average of surface deformation within a quarter annulus ($0 \leq \theta < \pi /2$). The lines indicate the loci of deformation maxima (orange), minima (blue)  and zero-crossing (green). These lines highlight that the deformation involves a decay in amplitude as well as a broadening of the lateral width of the deformation profile. Note that during this process, the position of the zero crossing (green line) remains approximately constant.\\

\subsection{Numerical simulation}

We perform numerical simulations in order to study the relaxation process of the viscous thin films. The relaxation process is modelled using lubrication theory {\color{blue} \citep{reynolds1886,oron1997long}}. Lubrication theory relies on the following conditions, which are indeed satisfied in the experiment, namely, (i) viscous forces in the film dominate over inertial forces ($Re_{f}$ $\sim$ $10^{-2}$ $\ll$ $1$) and (ii) deformation amplitudes in vertical direction are much lower than the characteristic lateral length scale ($\delta$/$\lambda$ $\sim$ $10^{-2}$ $\ll$ $1$). As boundary conditions we consider the free surface to be in contact with a homogenous gas pressure, as is the case after rebound, while there is a no-slip boundary condition at the substrate. The corresponding lubrication (thin film) equation reads  {\color{blue} \citep{oron1997long}}:
\begin{equation}
\partial_{t} h + \frac{\gamma_{_{f}}}{3 \eta_{_{f}}} \vec{\nabla} \cdot \left\{ h^{3} \vec{\nabla} \left( \vec{\nabla}^{2}h\right) \right\} = 0,
\label{eqn:equation00}
\end{equation}
where $h(x,y,t)$ is the vertical distance between the solid substrate and the free surface and $\vec{\nabla}$ is the two-dimensional gradient operator in the x-y plane. We perform a nondimensionalisation of \eqref{eqn:equation00} by $h = h_{f} H$, $r = h_{f} R$ and $t = (3 \eta_{f} h_{f} \gamma_{f}^{-1}) T$, where $h_{f}$ is the initial film thickness. In the following, we will study the relaxation process in an axisymmetric geometry, i.e. $H=H(R,T)$ such that \eqref{eqn:equation00} becomes :
\begin{equation}
\begin{aligned}
\partial_{T} H + \frac{1}{R} \partial_{R} \left[ R H^{3} \left( \partial_{R}^{3} H + \frac{1}{R} \partial_{R}^{2} H - \frac{1}{R^{2}} \partial_{R} H \right) \right] = 0.
\end{aligned}
\label{eqn:equation01}
\end{equation}

\begin{figure}
\begin{center}
\adjustbox{trim={0.04\width} {0.00\height} {0.06\width} {0.03\height}, clip}%
{\includegraphics[width=1.00\textwidth]{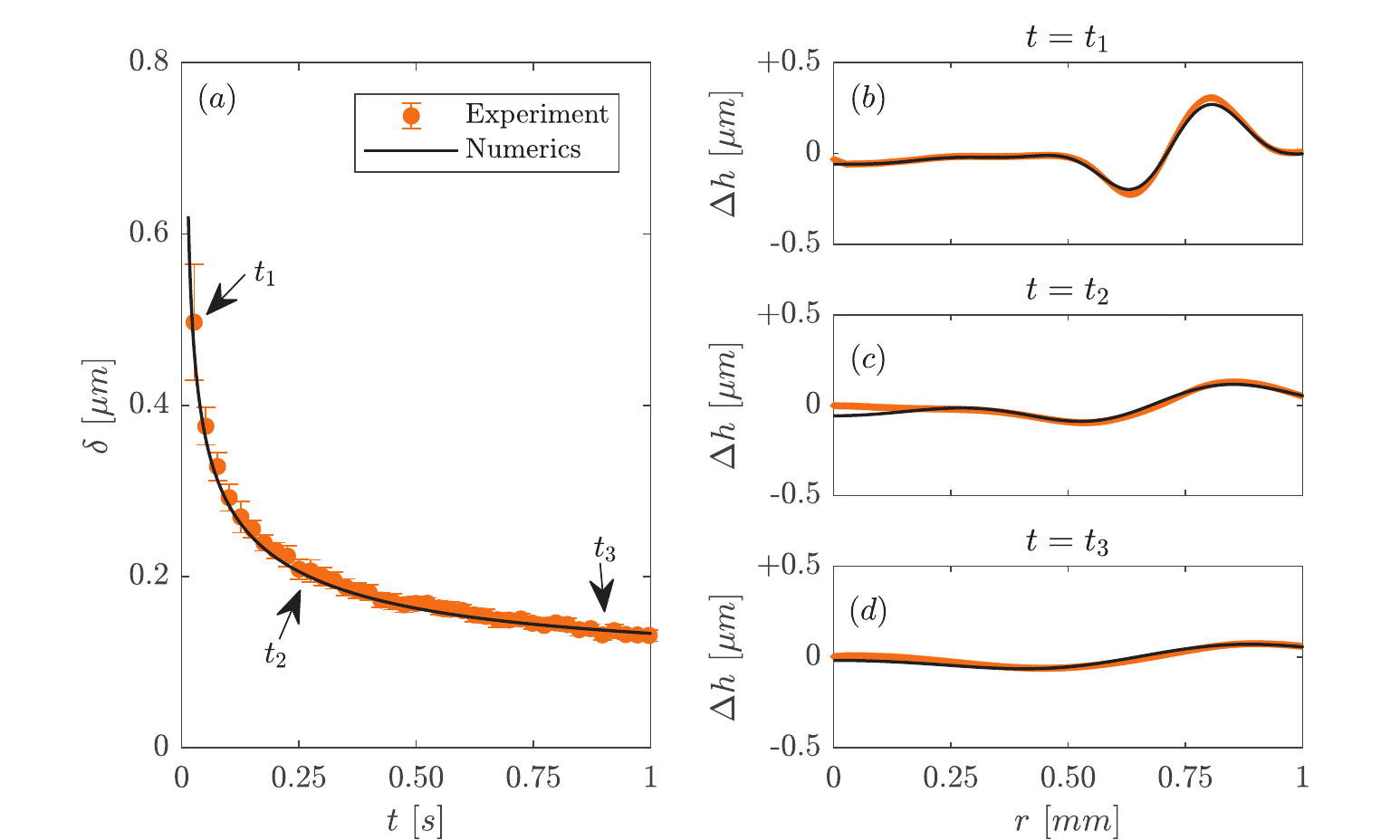}}
\end{center}
\caption{$(a)$ Comparison of amplitude decay between experiment and numerics. The mean and errorbar values of $\delta$ are obtained by binning every 25 datapoints. $(b-d)$ Comparison of film deformation between experiment and numerics at $t_{1}$ $=$ $14$ $ms$, $t_{2}$ $=$ $252$ $ms$ and $t_{3}$ $=$ $966$ $ms$. The control parameters are $We = 0.38$, $h^{*} = 0.01$, and $\eta^{*} = 52$.}
\label{fig:figure08}
\end{figure}

The asymptotics of \eqref{eqn:equation01} has also been studied by {\color{blue} \citet{salez2012numerical}}. Importantly, \eqref{eqn:equation01} is devoid of any free parameters. To compare to experiment, we perform numerical simulation of film relaxation using a finite element method and a second-order implicit Runge-Kutta time-stepping scheme (implemented using the framework {\sc{dune-pdelab}} by {\color{blue} \citet{bastian2008genericI, bastian2008genericII, bastian2010generic}}). The deformation profile at $t = 0$ is taken from the experiment and used as an initial condition, and subsequently the film profile is evolved via numerical integration of \eqref{eqn:equation01}. A direct comparison between the experiments and the lubrication theory is given in figure \ref{fig:figure08} without any adjustable parameters. Figure \ref{fig:figure08}a shows the amplitude $\delta$ (defined as the difference between maximum and minimum of $\Delta h$) as a function of time, while figure \ref{fig:figure08}b shows the deformation profiles $\Delta h = h(r,t) - h_{_{f}}$ at different times. The comparison exhibits very good agreement between the experiment and the numerical calculations, demonstrating the success of the lubrication approximation to describe the experimentally observed relaxation.

\subsection{Theoretical analysis}
Now we turn to a detailed theoretical analysis of the relaxation process, from which we will establish the general scaling laws of the relaxation. To do this, we reduce the lubrication equation to a one-dimensional geometry $h = h(x,t)$. Here, we use $X = x/h_{f}$ analogous to $R = r/h_{f}$. The rationale behind choosing a 1D lubrication equation is that the initial deformations are far from the impact center (cf. figure \ref{fig:figure04}). This is further quantified in figure \ref{fig:figure07}, where the ``width" of the profile is initially an order of magnitude smaller than the location  of the zero crossing. Therefore the axisymmetric relaxation and a one-dimensional relaxation will yield very similar results -- at least until the deformations approach the impact center and the azimuthal contributions become important. 

The one-dimensional lubrication equation reads

\begin{equation}
\begin{aligned}
\partial_{T} H + \partial_{X} \left( H^{3} \partial_{X}^{3} H \right) = 0.
\end{aligned}
\label{eqn:equation02}
\end{equation}
To further simplify the analysis, we use the fact seen in the experiments that the deformation amplitudes are small in comparison to initial film thicknesses ($\delta/h_{f} \sim 0.05 \ll 1$). This allows us to linearise \eqref{eqn:equation02} employing the variable transformation $Z = h/h_{f} - 1 = H - 1$ where $|Z| \ll 1$. The linearised 1D lubrication equation then reads
\begin{equation}
\begin{aligned}
\partial_{T} Z + \partial_{X}^{4} Z = 0.
\end{aligned}
\label{eqn:equation03}
\end{equation}
The relaxation of localised thin film perturbations described by \eqref{eqn:equation03} was analysed in great detail by {\color{blue} \citet{salez2012capillary, mcgraw2012self, baumchen2013relaxation, backholm2014capillary, benzaquen2013intermediate, benzaquen2014approach, benzaquen2015symmetry, bertin2020symmetrization}}. They obtained the long time asymptotic solution in terms of a moment expansion,

\begin{equation}
\begin{aligned}
Z (X,T) =  \frac{\mathcal{M}_{0} \phi_{0} (U)}{T^{1/4}} + \frac{\mathcal{M}_{1} \phi_{1} (U)}{T^{2/4}} + \frac{1}{2} \frac{\mathcal{M}_{2} \phi_{2} (U)}{T^{3/4}} + \, \dots ,
\end{aligned}
\label{eqn:equation04}
\end{equation}
which involves the similarity variable
\begin{equation}
\begin{aligned}
U = X T^{-1/4} ,
\end{aligned}
\label{eqn:equation05}
\end{equation}
and similarity functions $\phi_{n} (U)$ that can be determined analytically {\color{blue} \citep{benzaquen2013intermediate, benzaquen2014approach, benzaquen2015symmetry}}. The amplitudes $\mathcal{M}_{n}$ appearing in \eqref{eqn:equation04} can be determined from the initial condition $Z_0(X)=Z(X,0)$, by computing the moments 
\begin{equation}
\begin{aligned}
\mathcal{M}_{n} & = & \int_{}^{} \xi^{n} \, Z_{0}(\xi) \, d \xi, \qquad n=0,1,2,\,\ldots
\end{aligned}
\label{eqn:equation06}
\end{equation}

\begin{figure}
\centering
\includegraphics[width=\textwidth]{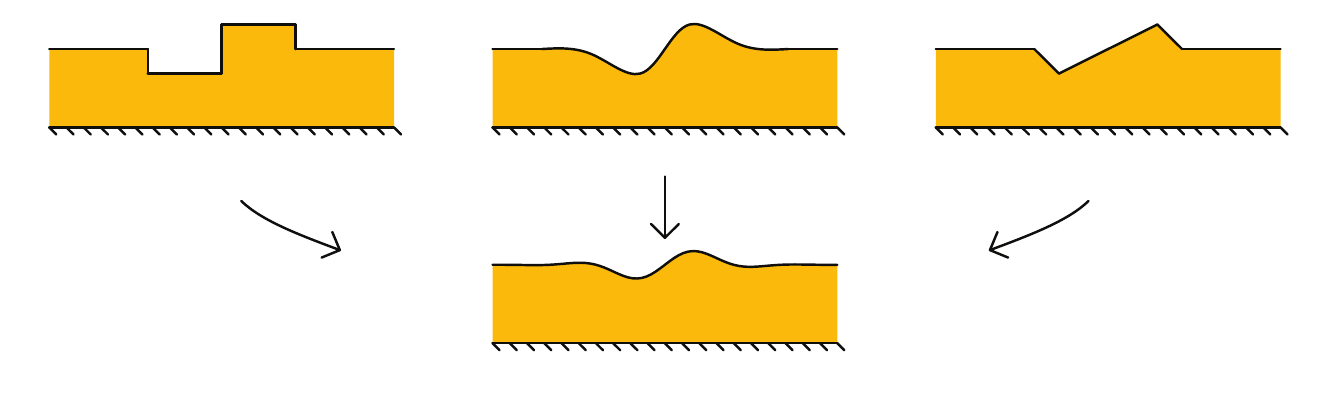}
\caption{Schematic illustrating of the approach to the attractor function $\phi_{1}(U)$. The initial deformations are zero volume perturbations having moments $\mathcal{M}_{0} = 0$ and $\mathcal{M}_{1} \neq 0$. Adapted from {\color{blue} \citet{benzaquen2014approach, benzaquen2015symmetry}}.}
\label{fig:figure09}
\end{figure}

It is clear from \eqref{eqn:equation04} and the similarity variable (\ref{eqn:equation05}) that the width $\lambda$ of the profile follows a universal scaling of the form $\lambda \sim T^{1/4}$. The decay of the amplitude $\delta$ is more subtle, since each term in \eqref{eqn:equation04} decays differently, as $\delta \sim T^{-(n+1)/4}$ for the $n$th moment. At late times, the solution $Z (X,T)$ thus converges towards the lowest order term with a non-zero moment. Generically, for $\mathcal M_0\neq 0$, the amplitude will therefore decay as $T^{-1/4}$. In our case, however, the perturbation originates from an initially flat film, and by incompressibility of the layer, the perturbation is thus expected to have a vanishing volume, i.e. $\mathcal{M}_{0} = 0$. In the present context, the lowest order moment is therefore expected to be $\mathcal{M}_{1} \neq 0$. In this scenario, the scaling law will be $\delta \sim T^{-1/2}$, while the solution $Z (X,T)$ should  converge to $\phi_{1}(U)$ for a zero volume perturbation. A schematic depiction of the approach to the $\phi_{1}(U)$ attractor is shown in figure \ref{fig:figure09}.

\begin{figure}[H]
\centering
\includegraphics[width=1.00\textwidth]{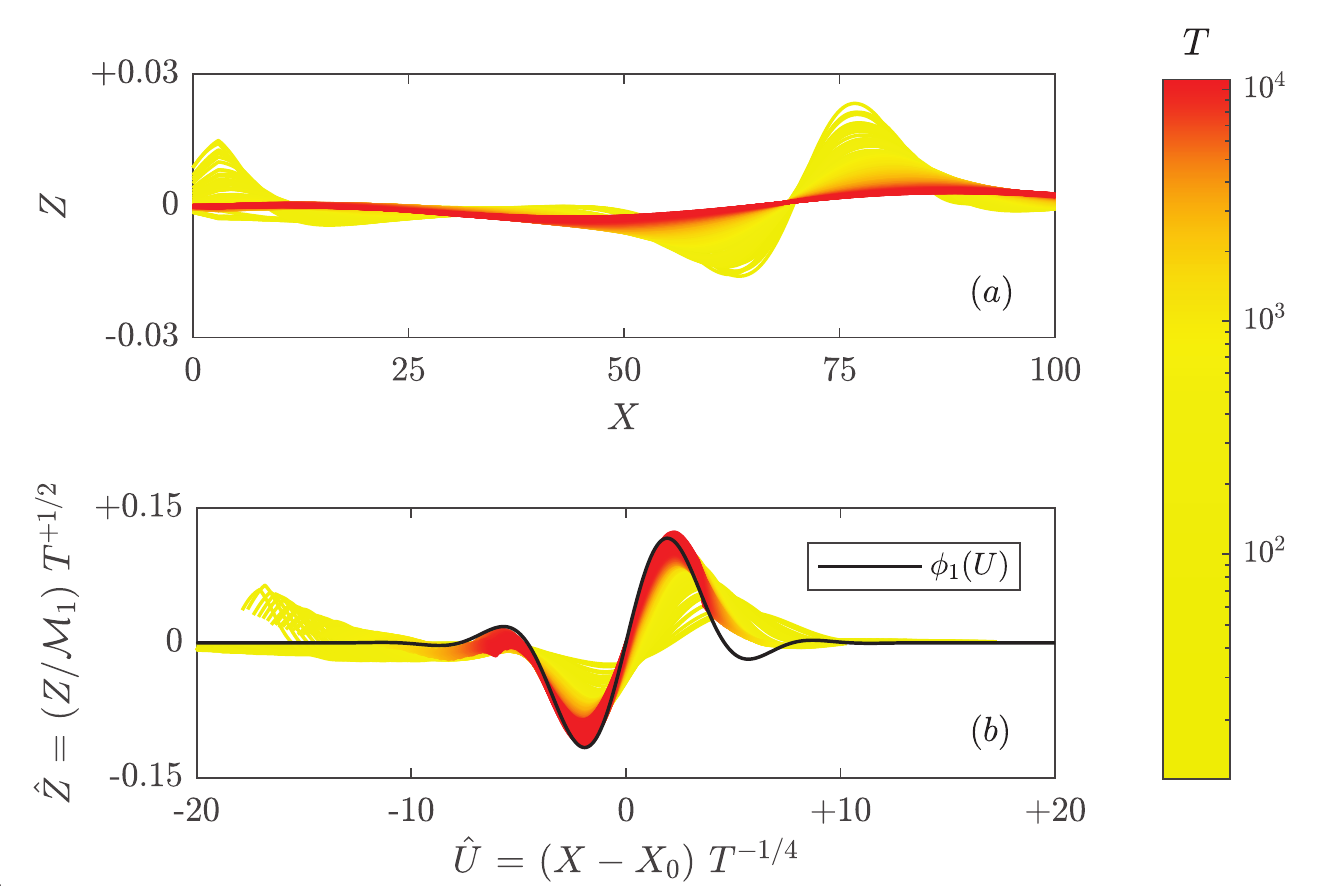}
\caption{(a) Time evolution of the normalized deformation profiles $Z$ vs $X$. A secondary deformation occurs at $T \approx 160$ near the impact center due to the next impact process (cf. figure \ref{fig:figure07}). (b) Time evolution of the scaled normalized deformation profiles $\hat{Z}$ vs $\hat{U}$. The self-similar attractor function $\phi_{1}(U)$ is plotted as a black line. Here, $X_{0} \approx 65$ and $\mathcal{M}_{1} \approx 3.2$. The scaled and rescaled deformations are color coded with time; yellow to red as time increases. The control parameters are $We = 0.38$, $h^{*} = 0.01$, and $\eta^{*} = 98$.}
\label{fig:figure10}
\end{figure}

To verify this scenario, we turn to an exemplary initial deformation profile of a $98$ $mPa.s,$ $10$ $\mu m$ film, and probe the subsequent relaxation. For the specific example, the two lowest order moments are determined as $\mathcal{M}_{0} \approx 6.4 \times 10^{-2}$ and  $\mathcal{M}_{1} \approx 3.2$. The very small value of $\mathcal{M}_{0}$ is of the order of the experimental resolution (i.e. it corresponds to a typical $\Delta h \sim \pm 10$ $nm$), so that indeed the perturbation has a negligible volume. Figure \ref{fig:figure10} reveals that the relaxation is indeed governed by $\mathcal M_1$, and approaches the $\phi_{1}(U)$ self-similar attractor function (cf. Appendix \ref{sec:Similarity function}). The figure reports the scaled deformation profiles centred around the zero crossing location $X_{0} \approx 65$. The first row shows the deformation profile scaled with the initial film thickness. The second row shows the rescaled deformation profiles (vertical scale $\sim T^{-1/2}$ and horizontal scale $\sim T^{1/4}$). During late times, the rescaled deformation profiles clearly approach the attractor function $\phi_{1}(U)$, which is superimposed on the data. This excellent match confirms that, within the experimental times, the axisymmetric effects have not yet started to contribute and the theoretical analysis of the film relaxation process over a 1D geometry is sufficient to understand the relaxation process in the experiments.

\subsection{Width broadening and amplitude decay}

Finally, we will compare theoretical asymptotic scaling laws for the width broadening and amplitude decay with a large number of experiments, all attained for a drop impact velocity of $v_{_{w}} \approx 0.16$ $m/s$. We predict the scaling law for the width $\lambda$ quantitatively, based on the approach to the attractor function $\phi_{1}$. We formally define the half-width of the similarity function as $U_{1}^{*} = \text{arg max } |\phi_{1}(U)| \approx 1.924$, which is half the absolute distance between global maxima and global minima (cf. figure \ref{fig:figure10}). From \eqref{eqn:equation05}, we then find 

\begin{equation}
\frac{\lambda}{2h_f } \simeq  U_{1}^{*} T^{1/4},
\label{eqn:equation07}
\end{equation}
expressing the dimensionless (half) width of the decaying profiles. A practical problem arises when comparing to experiment: at $t=0$, the width takes on a finite value $\lambda_0$, so that it is initially incompatible with the from (\ref{eqn:equation07}). To resolve this, we follow  {\color{blue} \citet{benzaquen2015symmetry}} and define for each experiment a convergence time $T_{\lambda}$ through $\lambda_{0} / 2h_{_{f}} = U_{1}^{*} T_{\lambda}^{1/4}$ (cf. Appendix \ref{sec:Convergence times}). The physical meaning of $T_{\lambda}$ is that it provides a time at which the experiment should approach the asymptotics power law  \eqref{eqn:equation07}. Using this definition of $T_\lambda$, the scaling \eqref{eqn:equation07} then gives
\begin{equation}
\begin{aligned}
\lambda/\lambda_{0}  \simeq  \left( T/T_{\lambda}\right)^{1/4},
\end{aligned}
\label{eqn:equation08}
\end{equation}
which can be compared to experiments without adjustable parameters. 

\begin{figure}
\centering
\includegraphics[width=1.00\textwidth]{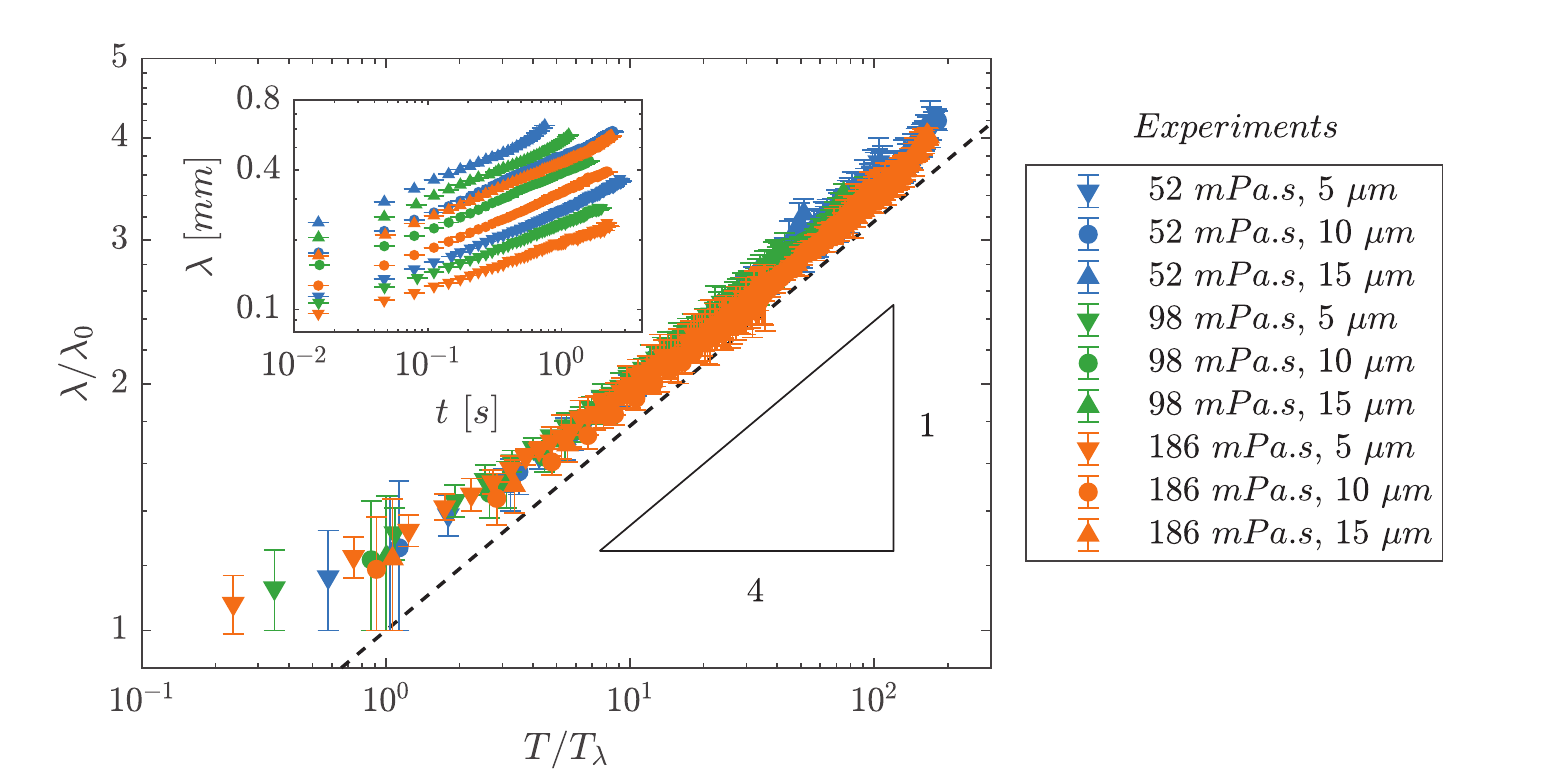}
\caption{Double logarithmic plot of $\lambda / \lambda_{0}$ vs $T / T_{\lambda}$. The mean and errorbar values of $\lambda/\lambda_{0}$ are obtained by binning every 25 datapoints. The black dashed line represents $\lambda/\lambda_{0} = \left( T / T_{\lambda} \right)^{1/4}$. The control parameters are $We = 0.38$, $h^{*} = 0.005-0.015$, and $\eta^{*} = 52-186$.}
\label{fig:figure11}
\end{figure}

In figure \ref{fig:figure11}, we show the temporal dependence $\lambda/\lambda_{0}$ vs $T/T_{\lambda}$ for different initial film thicknesses and viscosities. The black dashed line in the figure represents Eqn \eqref{eqn:equation08}. Clearly, all experimental data points seem to collapse onto a single master curve which is independent of the film properties used. Moreover, the master curve has a very good agreement with Eqn \eqref{eqn:equation08}. We remark that such a scaling  $\lambda \sim t^{1/4}$ is also seen in previous studies with viscous thin film configurations {\color{blue} \citep{salez2012capillary, mcgraw2012self, benzaquen2015symmetry, hack2018printing}}. Please note that during late times, some experiments show the width broadening to slowly deviate from the 1/4 scaling as seen in figure \ref{fig:figure11}. We suppose the deviations to come from the 1D to radial symmetry geometry transition and from the wave interactions coming from the second drop rebound which can contribute during late times.

A similar analysis is performed for the amplitude decay. We will once again make use of the self-similar attractor $\phi_{1} (U)$ for the relaxation process. Similar to the procedure outlined for the study of the width broadening, we employ Eqn \eqref{eqn:equation04} which implies,

\begin{equation}
\begin{aligned}
\frac{\delta}{2 h_{_{f}}} \simeq  \mathcal{M}_{1} \frac{ | \phi_{1}(U^*) | } {T^{1/2}},
\end{aligned}
\label{eqn:equation09}
\end{equation}
where $| \phi_{1}(U^*) |\approx 0.1164$ is the maximum of the similarity function (cf. figure \ref{fig:figure10}). To avoid the experimental issue that $\delta_0$, the amplitude at $t=0$, is finite, we once again determine for each experiment a convergence time $T_{\delta}$, using  $\delta_{0}/(2h_{_{f}}) =  \mathcal{M}_{1} | \phi_{1}(U^{*}) | /{T_{\delta}^{1/2}} $ (cf. Appendix \ref{sec:Convergence times}). Note that both $\delta_{0}$ and $\mathcal{M}_{1}$ will be different for each specific experiment, but both parameters can be determined independently. This finally gives that \eqref{eqn:equation09} can be written as

\begin{equation}
\begin{aligned} 
\delta/\delta_{0} \simeq (T/T_{\delta})^{-1/2}.
\end{aligned}
\label{eqn:equation10}
\end{equation}

\begin{figure}
\centering
\includegraphics[width=1.00\textwidth]{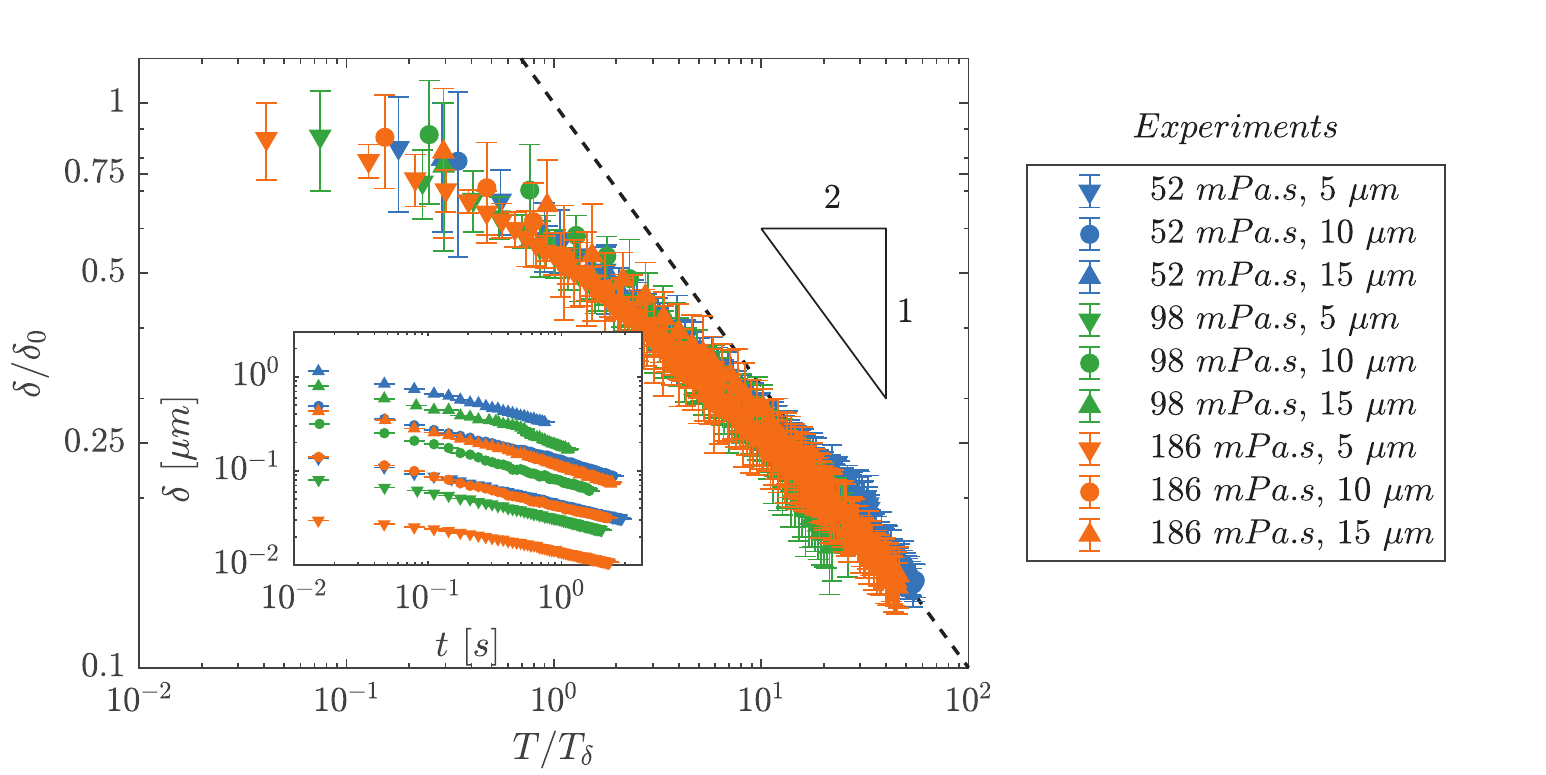}
\caption{Double logarithmic plot of $\delta / \delta_{0}$ vs $T / T_{\delta}$. The mean and errorbar values of $\delta/\delta_{0}$ are obtained by binning every 25 datapoints. The black dashed line represents $\delta / \delta_{0} = \left( T / T_{\delta} \right)^{-1/2}$. The control parameters are $We = 0.38$, $h^{*} = 0.005-0.015$, and $\eta^{*} = 52-186$.}
\label{fig:figure12}
\end{figure}

The result for $\delta/\delta_{0}$ vs $T/T_{\delta}$ for different film thicknesses and different viscosities are shown in figure \ref{fig:figure12}. The black dashed line in the figure represents Eqn \eqref{eqn:equation10}. While experiments exhibit a very good agreement with the numerical lubrication solution (cf. figure \ref{fig:figure08}), it is difficult to infer the scaling behaviour from individual realisations. However, it is clear that the rescaled amplitude plot of figure \ref{fig:figure12} is consistent with the predicted asymptotic decay. Indeed, the data seem to approach the 1/2 scaling law, as indicated by the dashed line.

\section{Conclusions and Outlook}
\label{sec:Conclusions and Outlook}

In this work, we have performed experiments of a water drop impacting a viscous thin oil film in an ambient air environment. The considered drop impact velocities are restricted to moderately low values $We \sim 1$ at which drops bounce on thin films, due to the air cushioning effect. Digital Holographic Microscopy was employed to measure the deformations of the free oil film surface that arise due to the bouncing of drops with an unprecedented precision, allowing for the one-to-one comparison with lubrication theory. 

We first investigated the deformations of the thin film immediately after rebound ($t = 0$) while varying the oil film thickness $h_{_{f}}$, oil viscosity $\eta_{_{f}}$, and the impact velocity $v_{_{w}}$ of the drop. We found that the deformation amplitude after the bounce $\delta_{0}$ varies quadratically with oil film thickness $\delta_{0} \sim h_{_{f}}^{2}$ and inversely with the oil viscosity $\delta_{0} \sim \eta_{_{f}}^{-1}$. When increasing the impact speed from $ v_{_{w}}\approx0.16~ m/s$ to $v_{_{w}}\approx0.37~ m/s$, the deformations in the thin film change even qualitatively: While at lower speeds, a single annular wavy deformation is found, at higher speeds the radial profiles exhibit two of such depressions and peak at two different radii.

In the second part of the manuscript, we have detailed the relaxation process of the viscous thin films when the drop is far away from the free oil surface after the bounce. Numerical calculations based on lubrication theory using the experimental deformations at $t = 0$ as initial condition show an excellent match with the experimentally measured evolution of the deformation profiles.
Furthermore, we have successfully employed a theoretical analysis developed by {\color{blue} \citet{benzaquen2013intermediate,benzaquen2014approach,benzaquen2015symmetry}} to obtain analytical results describing the relaxation process: Taking advantage of the fact that the deformations approach a universal self-similar attractor, at late times of the relaxation process the decay of the amplitude and the growth of the width of the deformations can be described without any free parameters. This allows us to collapse the corresponding experimental curves for all different thin film properties investigated.

Measuring the deformations of the falling drop and the viscous thin film simultaneously has proven challenging in the previous literature {\color{blue} \citep{lo2017mechanism}} and in the present experiments. However it is worthwhile to investigate the dynamics of the coupled system as it would allow valuable insight in the deformation mechanism.
To resolve both the drop-air and the oil-air interface, we therefore plan to combine the color interferometry technique by {\color{blue} \citet{van2012direct}}, which can be used to extract the narrow air profiles, with the DHM technique described in the present manuscript.
In order to understand the film deformations theoretically and quantify possible influences on the macroscopic drop dynamics, the macroscopic impact dynamics have to be coupled to a two layer lubrication model for the air layer and the lubrication layer. Very recently, such a lubrication approach has been pursued by {\color{blue} \citet{duchemin2020dimple}} for the coalescence of a drop with an underlying thin film and when combining color interferometry with DHM it can be tested against controlled experiments.

\section*{Acknowledgements}
This project has received funding from the European Union's Horizon 2020 research and innovation programme under the Marie Sklodowska-Curie grant agreement No 722497 and by the Max Planck Center Twente for Complex Fluid Dynamics. Kirsten Harth acknowledges funding from the German Science Foundation DFG within grants HA8467/1 and HA8467/2-1. We thank Gert-Wim Bruggert for constructing the impact setup, Guillaume Lajoinie for the insightful comments on the DHM results, Michiel Hack for aiding with the 3D image graphics and Mariia Zhakharova for fabricating the PDMS gels.

\section*{Declaration of interests}
The authors report no conflict of interest.

\appendix

\section{Measuring thin film deformations using DHM}
\label{sec:Measuring thin film deformations using DHM}

\begin{figure}[H]
\centering
\begin{subfigure}[b]{0.49\textwidth}
\centering
\includegraphics[width=0.875\textwidth]{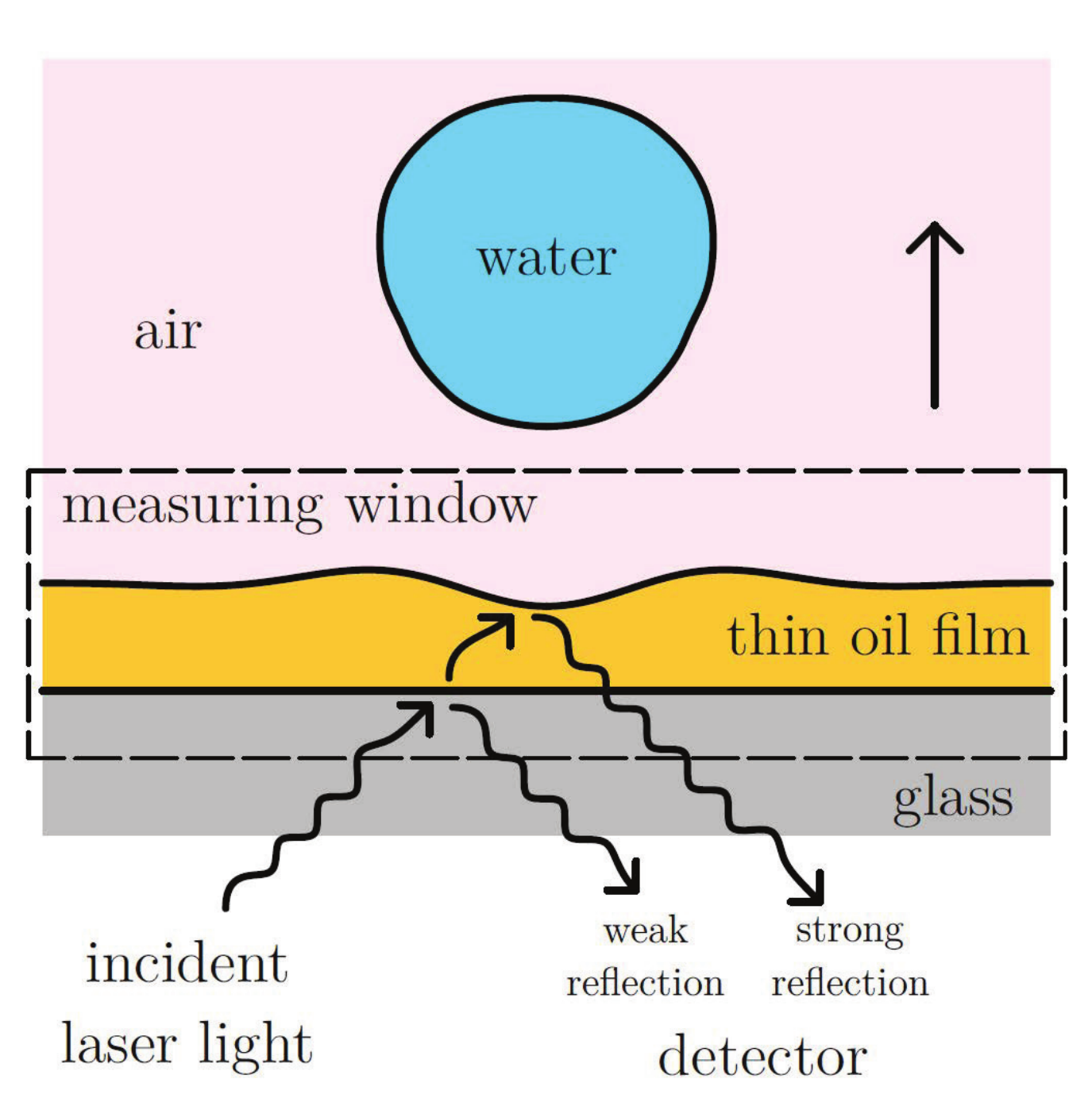}
\caption{}
\end{subfigure}
\begin{subfigure}[b]{0.49\textwidth}
\centering
\includegraphics[width=\textwidth]{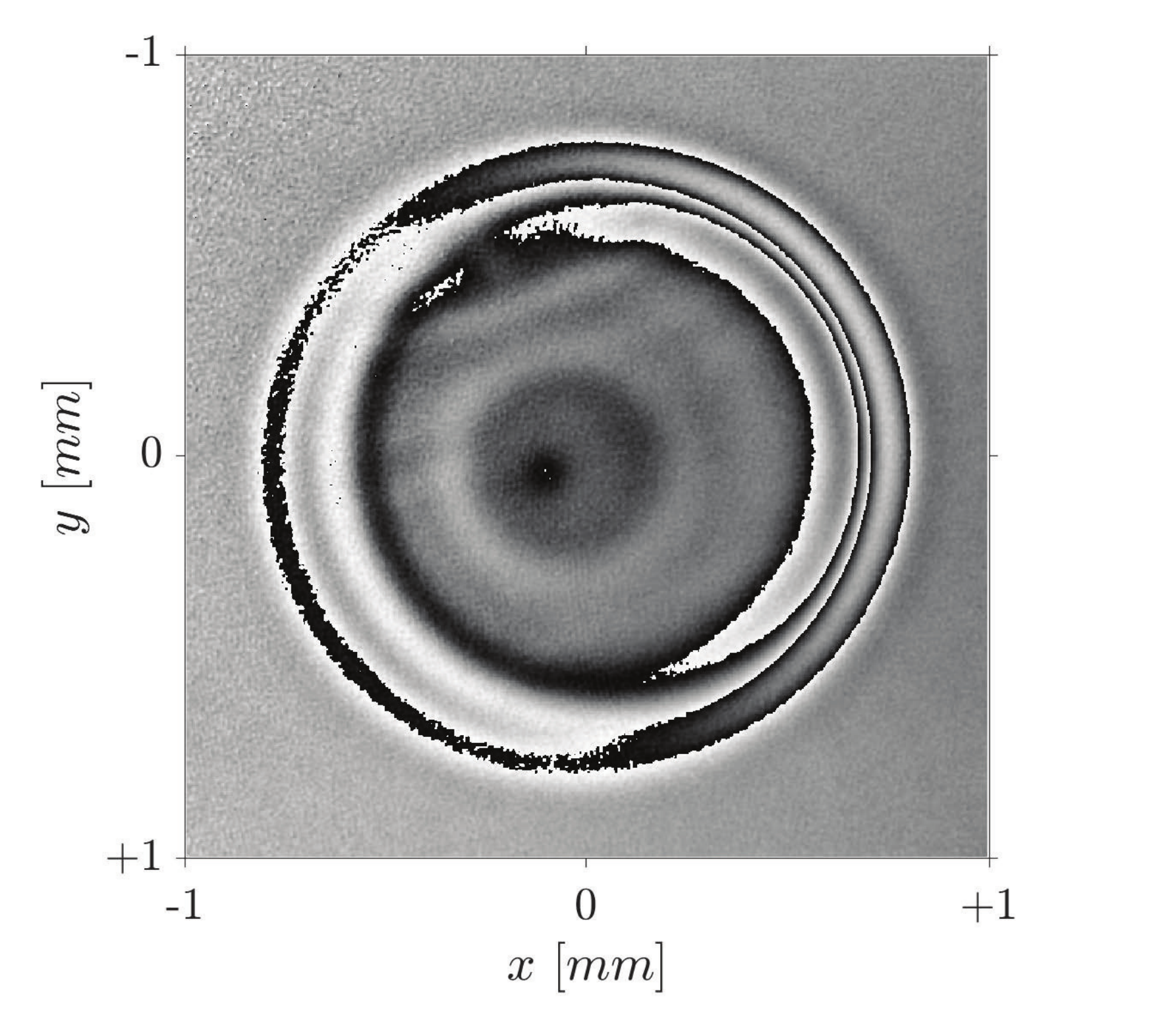}
\caption{}
\end{subfigure}
\caption{(a) Schematic diagram at $t = 0$ when the drop has just left the DHM measuring window after the bounce. (b) Phase image of the oil-air deformation at $t=0$. The control parameters are $We = 0.38$, $h^{*} = 0.01$, and $\eta^{*} = 98$.}
\label{fig:figure13}
\end{figure}

Figure \ref{fig:figure13}a shows a schematic diagram where the drop has just moved out of the DHM measuring window after the bounce. The height of the measuring window is roughly $200$ $\mu m$. Here, we define $t = 0$ as the first instance when the DHM measuring window is devoid of the water-air interface allowing for a clean measurement of the oil-air deformation. The object beam (consisting of reflection wavefronts from the measuring window) and the reference beam in the DHM interfere to form the holographic pattern recorded by the DHM camera.\\

\begin{figure}[H]
\centering
\includegraphics[width=1.00\textwidth]{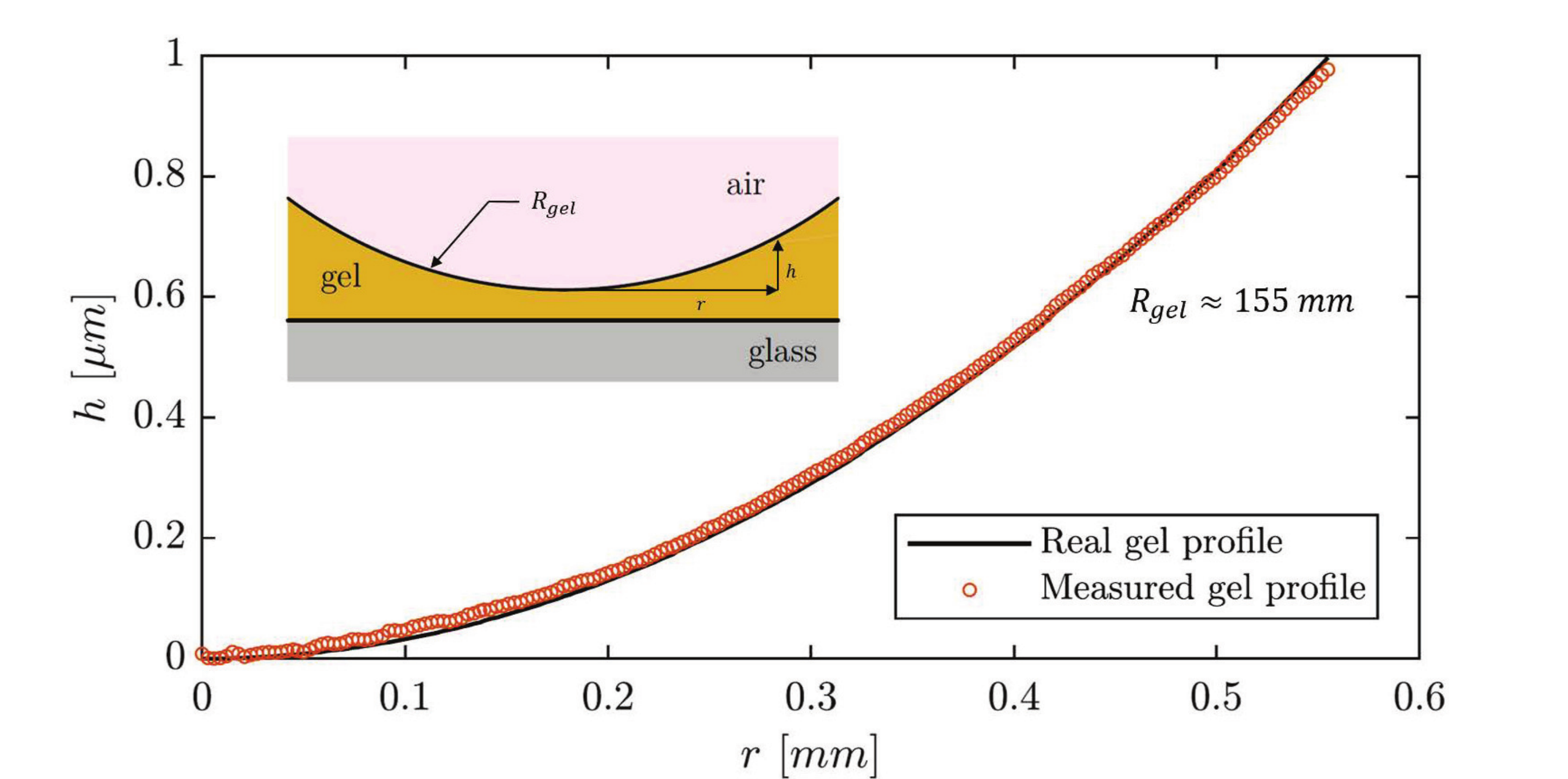}
\caption{Comparison of thin PDMS gel profiles of a known curvature.}
\label{fig:figure14}
\end{figure}

The recorded holograms are converted into their constituent intensity and phase information, wherein the phase information is used to reconstruct the height information. Figure \ref{fig:figure13}b shows an examplary phase image at time $t = 0$. The reconstructed height profile of this phase image is shown in figure \ref{fig:figure03}a.\\

To check whether the weak reflection of the glass-oil interface affects the measurements of the oil-air deformation in the bounce experiments, we perform a simple calibration experiment. The calibration experiment is performed to measure the (known) PDMS gel-air profile through glass. Figure \ref{fig:figure14} shows the comparison of the real and the measured PDMS gel-air deformation. We confirm that the flat glass-PDMS gel interface (and similarly the flat glass-oil film interface) does not affect the measurements of the thin film deformations. A vertical resolution of around $20$ $nm$ is obtained from the calibration plot. It is important to note that the vertical resolution can be well below $20$ $nm$. This only a conservative estimate since the real PDMS gel profile can suffer from small tilt issues and from surface roughness during its fabrication which are not accounted for here.\\

\section{Similarity function $\phi_{1}(U)$}
\label{sec:Similarity function}

In principle the similarity functions $\phi_{n} (U)$ are completely covered in the works of {\color{blue} \citet{benzaquen2013intermediate,benzaquen2014approach,benzaquen2015symmetry}}, but here we provide the expression for the similarity function $\phi_{1} (U)$ as a reference. The expression for $\phi_{1} (U)$ is given in Eqn \eqref{eqn:equation11} reads
\begin{equation}
\begin{aligned}
 \phi_{1} (U) = + & \frac{U}{4 \pi} \Gamma \left( \frac{3}{4} \right) {}_{0}H_{2} \left[ ; \left\{ \frac{5}{4}, \frac{3}{2}\right\}; \frac{U^{4}}{256} \right] \\
 - &\frac{U^{3}}{24 \pi} \Gamma \left( \frac{5}{4} \right) {}_{0}H_{2} \left[ ; \left\{ \frac{3}{2}, \frac{7}{4}\right\}; \frac{U^{4}}{256} \right] \\
 + &\frac{U^{5}}{960 \pi} \Gamma \left( \frac{3}{4} \right) {}_{0}H_{2} \left[ ; \left\{ \frac{9}{4}, \frac{5}{2}\right\}; \frac{U^{4}}{256} \right].
 \end{aligned}
\label{eqn:equation11}
\end{equation}
Here $\Gamma$ and ${}_{0}H_{2}$ are the gamma function and the (0,2)-hypergeometric function respectively. The (0,2)-hypergeometric function is defined in Eqn \eqref{eqn:equation12} which reads
\begin{equation}
{}_{0}H_{2} \left[ ; \left\{ a, b \right\}; w \right] = \sum_{n=0}^{\infty} \frac{1}{(a)_{n}\,(b)_{n}} \frac{w^{n}}{n!}.
\label{eqn:equation12}
\end{equation}
Here $(.)_{n}$ is the Pochhammer notation for the rising factorial.

\section{Convergence times $T_{\lambda}$ and $T_{\delta}$}
\label{sec:Convergence times}

The expressions for the convergence times $T_{\lambda}$ and $T_{\delta}$ are given in Eqn \eqref{eqn:equation13} reads,

\begin{equation}
\begin{aligned}
T_{\lambda} = \left( \frac{\lambda_{0}}{2 h_{_{f}} U_{1}^{*} } \right)^{4} \qquad \& \qquad T_{\delta} = \left( \frac{2 h_{_{f}} \mathcal{M}_{1} |\phi_{1} (U^{*})|} {\delta_{0}} \right)^{2}
\end{aligned}
\label{eqn:equation13}
\end{equation}

Here, $U_{1}^{*} \approx 1.924$ and $|\phi_{1} (U^{*})| \approx 0.1164$ are constants. Table \ref{tab:table02} shows the typical convergence time values obtained from the experiments.

\begin{center}
\begin{table}[H]
\begin{subtable}[t]{0.49\textwidth}
	\begin{center}
  	\begin{tabular}{|l|c|c|c|}
  	\cline{1-4}
  	              & $5$ $\mu m$ & $10$ $\mu m$ & $15$ $\mu m$\\\cline{1-4}
  	$52$ $mPa.s$  & $672$ & $173$  & $125$\\
	$98$ $mPa.s$  & $594$ & $121$  & $69$\\
	$186$ $mPa.s$ & $459$ & $60$   & $34$\\\cline{1-4}
\end{tabular}
\caption{$T_{\lambda}$}
\label{tab:table02a}
\end{center}
\end{subtable}
\begin{subtable}[t]{0.49\textwidth}
	\begin{center}
  	\begin{tabular}{|l|c|c|c|}
  	\cline{1-4}
  	              & $5$ $\mu m$ & $10$ $\mu m$ & $15$ $\mu m$\\\cline{1-4}
  	$52$ $mPa.s$  & $2187$ & $567$  & $449$\\
	$98$ $mPa.s$  & $2778$ & $421$  & $235$\\
	$186$ $mPa.s$ & $2648$ & $357$  & $124$\\\cline{1-4}
\end{tabular}
\caption{$T_{\delta}$}
\label{tab:table02b}
\end{center}
\end{subtable}
\caption{Convergence times $T_{\lambda}$ and $T_{\delta}$ values for varying film thickness and film viscosity. The control parameters are $We = 0.38$, $h^{*} = 0.005-0.015$, and $\eta^{*} = 52-186$.}
\label{tab:table02}
\end{table}
\end{center}

\bibliographystyle{jfm}
\bibliography{relaxation}

\end{document}